\pdfoutput=1
\documentclass[reprint,
superscriptaddress,
 amsmath,amssymb,
pra,
]{revtex4-2}
\renewcommand{\Re}{\operatorname{Re}}
\renewcommand{\Im}{\operatorname{Im}}
\usepackage{graphicx}
\usepackage{dcolumn}
\usepackage{bm}
\usepackage{comment}
\renewcommand{\selectlanguage}[1]{}
\newcommand{\tomega}{\overline{\omega}}
\newcommand{\omegan}{\tilde{\omega}}
\newcommand{\oG}{\overline{G}}
\newcommand{\tGamma}{\overline{\Gamma}}
\newcommand{\tgamma}{\overline{\gamma}}
\newcommand{\talpha}{\overline{\alpha}}

\newcommand{\Lpen}{\kappa}
\newcommand{\un}[1]{{\underline{#1}}}
\newcommand{\uun}[1]{{\underline{\underline{#1}}}}
\newcommand{\avec}{\un{a}}
\newcommand{\Omvec}{\uun{\Omega}}
\newcommand{\dvec}{\un{d}}
\usepackage[utf8]{inputenc}
\usepackage[T1]{fontenc}
\usepackage{mathptmx}
\usepackage{etoolbox}

\begin{document}

\preprint{APS/123-QED}

\title{Self-pulsing dynamics in  microscopic lasers with dispersive mirrors}

\author{Kristian Seegert}
\email{krsee@dtu.dk}
\affiliation{%
 Department of Electrical and Photonics Engineering, Technical University of Denmark, Building 345, 2800 Kgs. Lyngby, Denmark
}%
\affiliation{%
NanoPhoton - Center for Nanophotonics, Ørsteds Plads 345A, 2800 Kgs. Lyngby, Denmark
}%
\author{Mikkel Heuck}%
\author{Yi Yu}
\author{Jesper Mørk}
\affiliation{%
 Department of Electrical and Photonics Engineering, Technical University of Denmark, Building 345, 2800 Kgs. Lyngby, Denmark
}%
\affiliation{%
NanoPhoton - Center for Nanophotonics, Ørsteds Plads 345A, 2800 Kgs. Lyngby, Denmark
}%



\date{\today}

\begin{abstract}
We show that a passive dispersive reflector integrated into a semiconductor laser can be used to tailor the laser dynamics for the generation of ultrashort pulses as well as stable dual-mode lasing. We analyze the stability using a general model that applies to any laser with frequency-dependent mirror losses. Finally, we present a generalization of the Fano laser concept, which provides a flexible platform for tailoring the mirror dispersion for self-pulsing. In addition to functioning as a design guideline, our model also accounts for several results in the literature. 

\end{abstract}

\maketitle

\begin{figure}
    \includegraphics[width=1\linewidth]{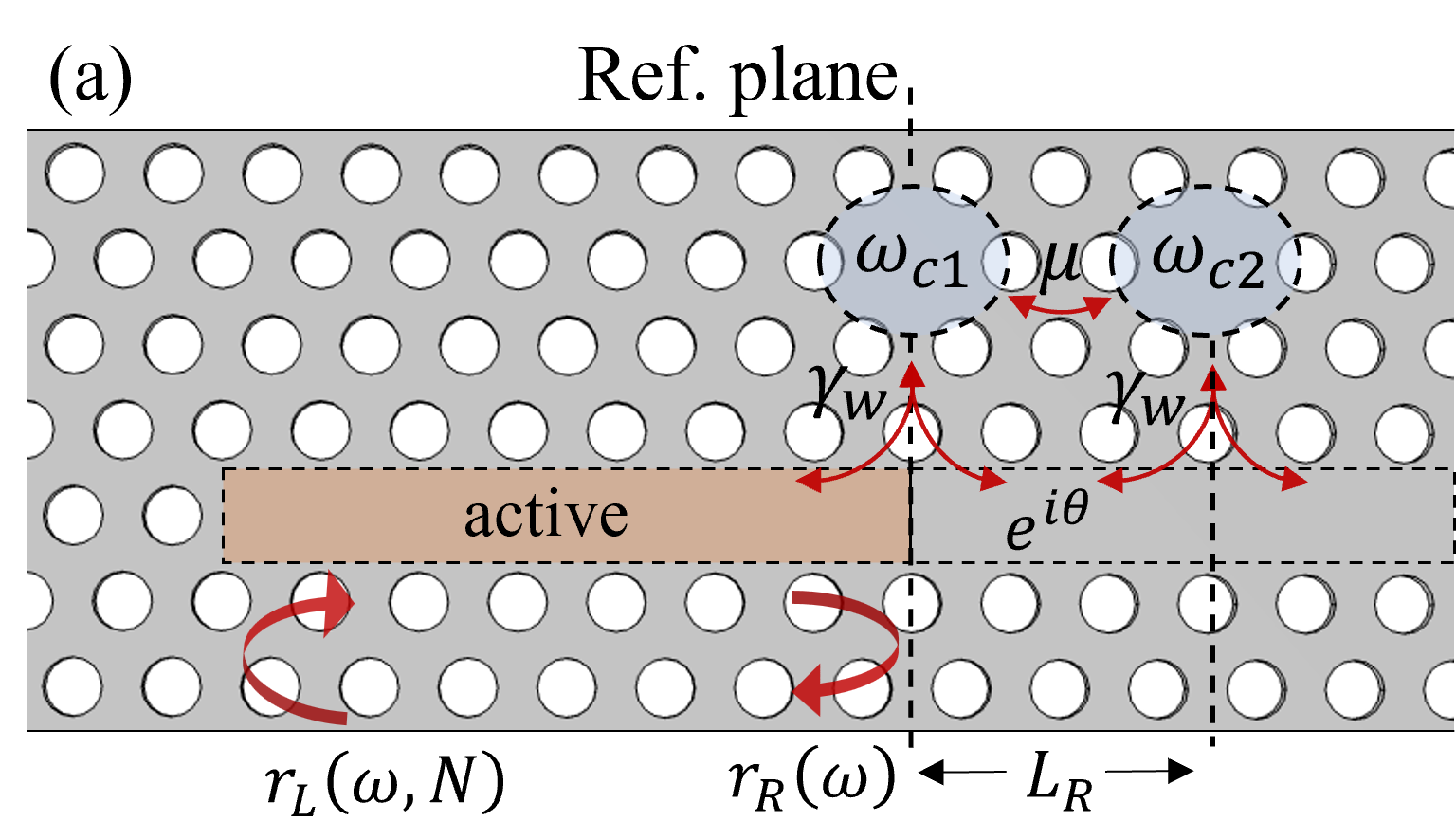}
    \includegraphics[width=\linewidth]{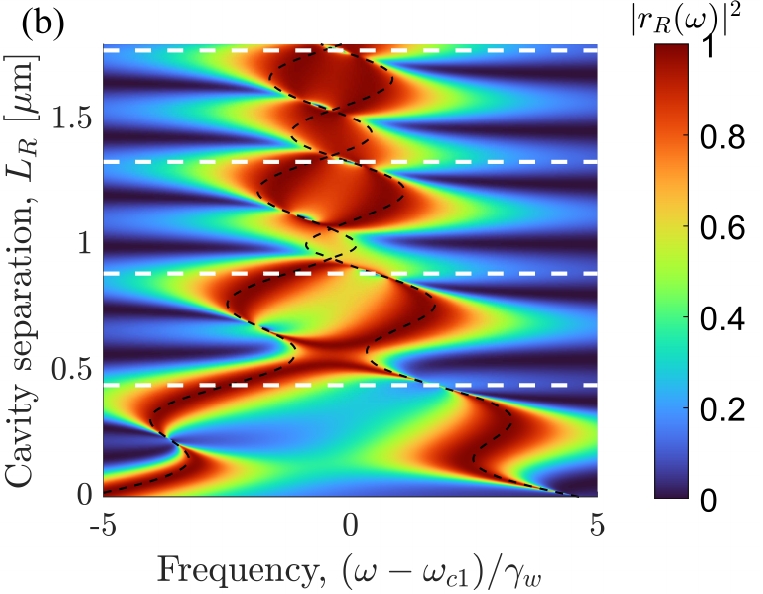}
     \caption{(a) Sketch of the coupled-cavity Fano laser. Reflection off of the left mirror and propagation through the active region with carrier density, $N$, is described by $r_L(\omega,N)$. The reflection from the two coupled cavities on the right is $r_R(\omega)$. The cavities, with resonances at $\omega_{c1}$ and $\omega_{c2}$ are directly coupled with a rate $\mu$ and indirectly through a waveguide of length $L_R$ and propagation constant $k$. (b) 2D reflectivity map of $|r_R(\omega)|^2$ (colour scale) vs frequency $\omega$ and cavity separation $L_R$. The coupling phase is $\theta=k L_R$, and the direct coupling between is modeled as $\mu = 5\gamma_w\exp(-L_R/\Lpen)$ where $\Lpen=0.5\mu$m. The white dashed lines denote where $\theta =0$ mod $2\pi$. The detuning between the cavities is set to $\Delta=-0.78\gamma_w$ ($\Delta/2\pi=$ -100 GHz), and the Q-factor related to coupling to the waveguide is set to $Q_w=750$ for both cavities. The intrinsic quality factor is set to $Q_i=10^5$.} 
     \label{fig:1}
\end{figure}
\section{Introduction}
 The generation of optical pulses plays a crucial role in many photonic technologies, including communications~\cite{hu_chip-based_2021}, spectroscopy~\cite{udem_optical_2002}, all-optical clock recovery~\cite{feiste_18_1994}, sensing~\cite{suh_microresonator_2016}, and LiDAR. In addition, excitable spiking nanolasers may act as "photonic neurons" in neuromorphic computing~\cite{pammi_micro-lasers_2020,shastri_photonics_2021}. 
Much effort has been put into reducing the size and increasing the energy efficiency of pulsed lasers~\cite{miller_attojoule_2017}. However, passive $Q$-switching in a microscopic laser was only demonstrated recently~\cite{yu_demonstration_2017,delmulle_self-pulsing_2022}. All current demonstrations of $Q$-switched nanolasers rely on the presence of an element in the laser cavity that exhibits saturable absorption, such that the laser favors operation in a pulsed rather than a continuous wave state. Saturable absorbers require control of the carrier lifetimes, which need to be shorter in the absorber section than in the gain section~\cite{tronciu_self-pulsation_2003}. This is done either by using different materials, reverse biasing the absorber section, or modifying the lifetimes in other ways, e.g., passivating the active section. 
 
An alternative way of generating self-pulsing relies on passive dispersive reflectors (PDRs)~\cite{wenzel_mechanisms_1996,rimoldi_damping_2022,ramunno_stability_2002,mak_linewidth_2019,bandelow_tailoring_2000,yu_theory_2022}. Compared to saturable absorbers, PDRs may allow a lower lasing threshold (there are no losses that first need to be saturated) and engineered output pulses. Another advantage of PDRs is that they are material-independent, relying only on the geometry of the design. 

 So far, modeling of lasers with dispersive mirrors has mainly been done by implementation-specific approaches, and this prevents general conclusions from being drawn about their possibilities and limitations.
 Here, we provide a unified description of how a dispersive mirror influences the dynamical properties of the laser and, in particular, its stability. For microscopic lasers, referring only to the local slope and curvature of the mirror reflection spectrum, we are able to predict the onset of self-pulsing.
 
We keep the details of the PDR general and instead focus on the inverse problem: what kinds of dynamic instabilities can arise, and how does the onset of instabilities relate to the PDR reflection spectrum? We show that a generalization of the Fano laser~\cite{mork_photonic_2014,mork_semiconductor_2019} provides a flexible platform for tailoring the mirror response. In particular, we demonstrate the possibility of generating short optical pulses (see section \ref{sec:dsq}), as well as stable dual-mode lasing corresponding to beating oscillations (see section \ref{sec:beat}) with a tunable beat-note frequency much smaller than the free spectral range of the cavity. As a key result, we derive a modified characteristic equation for the linearized system that explicitly takes into account the frequency-dependent mirror response. We derive a general expression for the relaxation oscillation frequency and damping rate that depends on the local shape of the reflectivity $r_R(\omega)$. This expression may be applied as a guideline for the design of the PDR. Furthermore, it provides a simple and general explanation of various results already presented in the literature regarding the impact of a frequency-dependent mirror on relaxation oscillations.

\subsection{The coupled-cavity Fano laser}
Our proposed generalization of the Fano laser is sketched in Fig. \ref{fig:1}a. It consists of a semi-open waveguide that is side-coupled to two nanocavities. The left mirror is broadband and formed by terminating the waveguide, while the right mirror is based on Fano interference between the nanocavities and the waveguide, making its reflectivity, $r_R(\omega)=|r_R(\omega)|\exp(i\phi(\omega))$, strongly frequency-dependent~\cite{fano_effects_1961}. Importantly, a buried heterostructure~\cite{matsuo_high-speed_2010,yu_ultra-coherent_2021} ensures that gain material only exists in the waveguide segment between the left mirror and the leftmost nanocavity. The nanocavities are thereby completely passive, with $r_R(\omega)$ being independent of the carrier density, $N$. 

The original Fano laser is based on a single side-coupled nanocavity~\cite{mork_photonic_2014,mork_semiconductor_2019}.
In the case of a single nanocavity, the transmission in the waveguide below the nanocavity will have two contributions corresponding to two different optical pathways; namely, direct propagation in the waveguide, and coupling to the nanocavity and back into the waveguide in the same direction. If the frequency of the incoming wave matches the resonance frequency of the nanocavity, destructive inference occurs between these two optical pathways, and the result is a narrow-band mirror with a Lorentzian reflection spectrum~\cite{fan_temporal_2003}. The conventional Fano laser has already shown many interesting properties and dynamics including the theoretical possibility of terahertz frequency modulation~\cite{mork_photonic_2014}, stability towards coherence collapse~\cite{rasmussen_suppression_2019}, self-pulsing (in the case where the active material extends into the nanocavity~\cite{yu_demonstration_2017}), ultra-narrow linewidth (in the case where the active material is confined to the waveguide section similar to Fig. \ref{fig:1})~\cite{yu_ultra-coherent_2021}, and the possibility to dynamically modulate the mirror losses~\cite{dong_cavity_2023}. In addition to functioning as narrow-band mirrors, Fano resonances also have interesting possible applications in all-optical switching, signal-processing, and frequency-conversion~\cite{bekele_plane_2019}.

In the present case, the addition of a second nanocavity to the original Fano laser~\cite{mork_photonic_2014} allows additional possibilities for engineering the mirror reflectivity, which can be analyzed by temporal coupled-mode theory~\cite{wonjoo_suh_temporal_2004}. Note that in addition to being placed on the same side of the waveguide as depicted in Fig.~\ref{fig:1}a, the cavities can also be placed on opposite sides as in Refs.~\cite{yan_controlling_2023,heuck_dual-resonances_2014}. The spectral response of the dispersive reflector is determined by the cavity detuning $\Delta = \omega_{c2}-\omega_{c1}$, the direct coupling rate $\mu$, and the indirect waveguide-coupling described by the phase accumulation $\theta=kL_R$, see Fig.~\ref{fig:1}a.

Figure~\ref{fig:1}b shows a 2D map of the reflectivity $|r_R(\omega)|^2$ versus frequency and nanocavity separation, $L_R$, for a detuning of $\Delta/2\pi = -100$ GHz. In order to represent the dependence on both the direct and indirect coupling in a straightforward manner, we use a simple model where the direct coupling falls off as $\mu =\mu_0 \exp(-L_R/\Lpen)$, where $\Lpen$ is some characteristic length. The white dashed lines show where $\theta=2\pi m$. 

We can understand the reflectivity spectrum in Fig.~\ref{fig:1}b as follows: The two cavities form two supermodes with complex frequencies $\omega_\pm$ which are hybridizations of the modes of the uncoupled cavities, and each supermode will be accompanied by a Lorentzian frequency dependence of the reflectivity. The real part of the resonance frequencies are shown as the black dashed lines. In our case, the complex supermode frequencies are given by 
\begin{equation}
    \omega_{\pm}=\frac{\omega_{c1}+\omega_{c2}}{2}-i(\gamma_w+\gamma_i)\pm \sqrt{\left(\frac{\Delta}{2}\right)^2+(\mu+i\gamma_we^{i\theta})^2},
\end{equation}
where $\gamma_w$ is the coupling rate between the nanocavities and the waveguide, and $\gamma_i$ represents intrinsic losses, which are assumed to be identical for the two cavities.
When the nanocavity separation $L_R$ is small, such that the direct coupling is strong ($\mu\gg \gamma_w$), the resonances are clearly split, but they start to overlap as the cavities get further apart. The reflection occurs via different optical paths, and interference between them causes variations in the positions and widths of the Lorentzian resonances. 

If the cavities have identical resonance frequencies ($\Delta=0$), the supermodes are even and odd with respect to reflections through a symmetry plane located between them. For $e^{i\theta}$ approaching 1, the outgoing waves of each cavity interfere constructively (destructively) for the odd (even) supermode leading to broadening (narrowing) of the associated resonance peak, making the spectrum highly asymmetric. This phenomenon of loss-splitting is referred to as \textit{dissipative coupling}, and exactly when 
$e^{i\theta}=1$ (white dashed lines in Fig.~\ref{fig:1}b), we get an example of a bound-state-in-the-continuum (BIC), which does not couple to the waveguide at all~\cite{hsu_bound_2016}. The same effect takes place for $e^{i\theta}=-1$, but with the broadening/narrowing being reversed. On the other hand, when $e^{i\theta}=\pm i$, the out-going waves are phase-shifted by $\pi/2$, which only affects the real part of the resonance frequencies, resulting in a symmetric spectrum. This case, where $\Im(\mu+i\gamma_we^{i\theta})=0$, is referred to as \textit{dispersive coupling}. The general considerations on the effect of dispersive and dissipative coupling also hold when the cavities are slightly detuned, although the supermodes will not be exactly even and odd.

While coupled mode theory has been shown to accurately account for the dispersive properties of systems composed of coupled waveguides and cavities~\cite{kristensen_theory_2017}, actual designs will, of course, need to rely on numerical calculations, using e.g. FDTD or finite-element calculations. 

Using dual-cavity reflectors, several works illustrate the large freedom one has to engineer the spectral shape of the reflection spectrum by modifying the geometry of the cavities, including their relative positions~\cite{chalcraft_mode_2011}, the potential barrier between the cavities~\cite{haddadi_photonic_2014}, and possibly adding blocking elements in the waveguide~\cite{yan_controlling_2023}. 
Finally, in addition to tuning the response through the designed geometry, it is also possible to dynamically modulate the cavities through nonlinear effects~\cite{yu_switching_2013,dong_cavity_2023}, or by electrodes that change the refractive index of the nanocavity through an applied electrical field.

The ability to manipulate the mirror's response opens up new avenues to design laser dynamics, as the wide range of reflection spectra may lead to very different dynamical regimes. Next, we turn to the question of how a frequency-dependent mirror affects the laser dynamics. After the general analysis, we provide two examples of applications: Self-Q-switching and dual-mode lasing.

\section{General Stability analysis}
\begin{figure}
    \centering
    \includegraphics[width=0.9\linewidth]{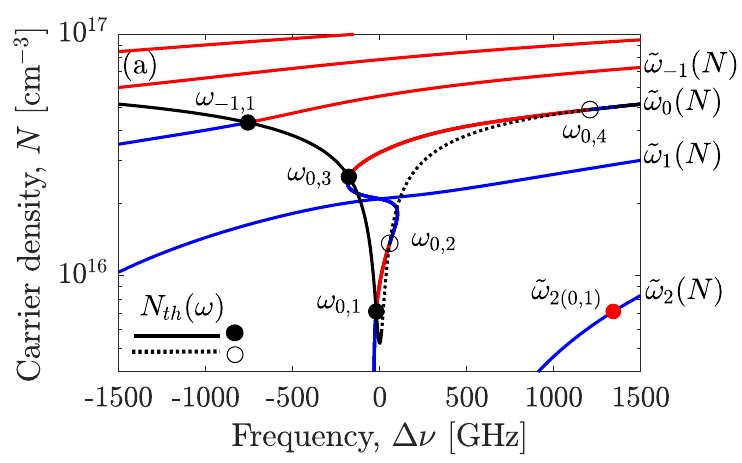}
    \includegraphics[width=0.9\linewidth]{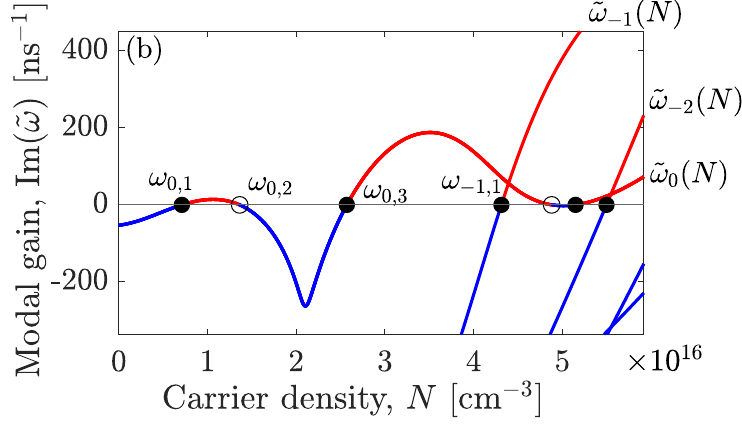}
    \caption{Illustration of the various concepts introduced in section II in the $(\Delta\nu,N)$-plane (a) and $(N,\Im(\omegan))$-plane (b), using an example of a Fano laser with a Lorentzian reflection spectrum. The frequency axis in (a) is shifted relative to the reflectivity peak $\Delta\nu = (\Re(\omegan)-\omega_c)/2\pi$. Colored curves are instantaneous modes $\omegan_n(N)$, with $\Im(\omegan)>0$ in red and $\Im(\omegan)<0$ in blue. Black dots are steady-state modes ($\omega_s,N_s$). Empty black circles are antimodes. Red dots are sidemodes of the lowest threshold steady-state mode. The black curve in (a) is $N_{th}(\omega)$, which is fully drawn when $\tGamma(\omega)>0$ and dotted when $\tGamma(\omega)<0$, following Eq. (\ref{eq:eff_parameters}).}
    \label{fig:modes}
\end{figure}\subsection{Modal backbone of lasers with dispersive mirrors}
The general analysis in this section applies to any laser that can be modeled as an effective Fabry-Perot laser with a dispersive mirror. In order to get a better understanding of the dynamics of lasers with dispersive mirrors, we will introduce the concepts of \textit{steady-state modes}, \textit{antimodes}, \textit{instantaneous modes}, and \textit{sidemodes}. Additionally, steady-state modes and antimodes are referred to collectively as \textit{steady-state solutions}.

We define the forward- and backward-propagating complex electric fields at a reference plane just left of the PDR (see Fig.~\ref{fig:1}a) as $E(\omega)$ and $E_-(\omega)$.
They are related by~\cite{detoma_complex_2005} \begin{align}
    \label{eq:Eplus}
    E(\omega)&=r_L(\omega,N)E_-(\omega)+F(\omega),\\
    \label{eq:Eminus}
    E_-(\omega)&= r_R(\omega) E(\omega),
\end{align}
where $F(\omega)$ is a term representing spontaneous emission and noise. The function $r_L(\omega,N)$ is given by \begin{equation}
    r_L(\omega,N)=r_1e^{2ik(\omega,N)L},
\end{equation} where $r_1$ is the reflectivity of the frequency-independent left mirror, $k(\omega,N)$ is the wavenumber, and $L$ is the length of the active section.

We define the \textit{instantaneous modes} $\omegan_n(N)$ as the complex solutions to the oscillation condition
\begin{equation}
    r_L(\omegan_n,N)r_R(\omegan_n)=1, \qquad \omegan_n\in\mathbb{C},
\end{equation}
where the carrier density is interpreted as a parameter. The instantaneous modes trace out branches of solutions in the complex frequency plane that depend parametrically on $N$, and the subscript "$n$" denotes a particular solution branch. The name "instantaneous modes" is taken from analogous concepts in Refs.~\cite{wenzel_mechanisms_1996,piprek_optoelectronic_2005}, and refers to the fact that they solve the oscillation condition for a fixed instantaneous carrier density, even though the carrier density is, in general, a function of time. 
The imaginary part of an instantaneous mode gives the effective net modal gain per unit time $\oG_n(N)=2\Im(\omegan_n(N))$, and we define the modal differential gain per unit time as $\oG_{Nn}(N)=\frac{d}{dN}\overline {G}_n(N)$. In this paper, "effective" parameters are characterized with an overline.

In Fig.~\ref{fig:modes}a the instantaneous modes are drawn as the red/blue curves in $(\Delta \nu,N)$-space, where $\Delta\nu = (\Re(\omegan)-\omega_c)/2\pi$, using the example of a single-cavity Fano laser with a Lorentzian reflection spectrum $r_R(\omega)\propto 1/(\omega-\omega_c-i\gamma)$~\cite{mork_photonic_2014}. Blue (red) means the mode is below (above) threshold, as illustrated in Fig.~\ref{fig:modes}b, which shows the modal gain $\oG_n(N)$ as a function of carrier density.

The points where $\omegan_n(N)$ become real-valued, such that $\oG_n(N)=0$, define the steady-state solutions ($\omega_s,N_s$) which satisfy the oscillation condition \begin{eqnarray}
\label{OC}
    r_L(\omega_s,N_s)r_R(\omega_s)=1, \qquad (\omega_s,N_s)\in \mathbb{R}^2.
\end{eqnarray}
In the time domain, the steady-state solutions correspond to continuous-wave (CW) operation at a certain frequency and carrier density. The subscript "$s$" here denotes a set of two indices, $s=(n,j)$, corresponding to the $j$'th steady-state solution belonging to the $n$'th instantaneous mode, $\omega_{n,j} = \omegan_n(N_{n,j})$. 

The steady-state solutions can further be divided into \textit{steady-state modes} and \textit{antimodes}, depending on whether the associated effective gain crosses zero in the positive ($\oG_{Nn}(N_s)>0$) or negative ($\oG_{Nn}(N_s)<0$) direction. In Figs.~\ref{fig:modes}a and~\ref{fig:modes}b, steady-state modes are marked with filled circles, while antimodes are marked with empty circles. Only the steady-state modes can be stable, while antimodes are always unstable and correspond to saddle-nodes~\cite{tromborg_mode_1997}.

As shown in Fig.~\ref{fig:modes}a, all steady-state solutions $(\omega_s,N_s)$ fall on the line $N_{th}(\omega)$, which solves the amplitude condition $|r_L(\omega,N_{th}(\omega))r_R(\omega)|=1$. The shape of $N_{th}(\omega)$ mimics the reflection spectrum $r_R(\omega)$, and is given by 
\begin{eqnarray}
    \Gamma g(N_{th}(\omega))=\alpha_i+\frac{1}{2L}\ln\left(\frac{1}{|r_1r_R(\omega)|^2}\right),
\end{eqnarray}
where $\Gamma$ is the confinement factor, $g(N)$ is the material gain, $\alpha_i$ represents intrinsic losses in the waveguide, and $r_1$ is the reflectivity of the left broadband mirror. The actual positions of steady-state modes are then given by the phase condition along the threshold carrier density curve, 
\begin{eqnarray}
\label{eq:phasecondition}
    \arg(r_R(\omega_s))+\arg(r_{L}(\omega_s,N_{th}(\omega_s)))=2\pi p+\phi_0,
\end{eqnarray}
where $p$ is an integer, and $\phi_0$ is a global phase representing the possible inclusion of some phase tuning mechanism. Numerically, looking for solutions to the phase condition along the threshold carrier density curve makes finding the steady-state solutions a simple task. The instantaneous modes can then be computed numerically with path-continuation starting from each steady-state solution.

Finally, we define the \emph{sidemodes} $\omegan_{ms}\equiv \omegan_{m}(N_s)$, which should be understood as the $m$'th sidemode of the steady-state solution $s=(n,j)$ where $m\neq n$. The sidemodes thus solve the oscillation condition at the carrier density level $N=N_s$ 
\begin{eqnarray}
r_L(\omegan_{ms},N_s)r_R(\omegan_{ms})=0,\qquad  m\neq n. 
\end{eqnarray} 
If the laser oscillates in the steady-state mode $s$, then the sidemodes are the other instantaneous modes simultaneously present in the system. 

Letting $R_p$ denote the pump rate and $\tau_s$ the carrier lifetime, then immediately above the threshold $R_p\gtrsim N_{s}/\tau_s$ of the steady-state mode $(\omega_{s},N_{s})$, its stability is determined by the positions of the sidemodes. If just one sidemode experiences gain, $\Im(\omegan_{ms})>0$, the steady-state mode $(\omega_{s},N_{s})$ is unstable; otherwise it is stable~\cite{tromborg_mode_1997}. 
In the absence of antimodes, which is the case for a regular Fabry-Perot laser, the instantaneous modes stay above threshold once they have been reached ($\Im(\omegan_n(N))>0$ for $N>N_{s}$). This implies that only the steady-state mode with the lowest threshold will not have one or more sidemodes experiencing gain. In the presence of antimodes, however, this is not necessarily the case. An example of this is $\omega_{0,1}$ and $\omega_{0,3}$ in Fig.~\ref{fig:modes}b, which are both stable at threshold, since all their respective sidemodes are below threshold, and this is possible due to the antimode $\omega_{0,2}$. 

Further, as the pump rate is increased above threshold, steady-state modes that are initially stable may become unstable due to, e.g., the presence of a weakly damped sidemode. Similarly, steady-state modes, which are initially unstable but only weakly suppressed, may become stable. These mode-coupling phenomena are sometimes referred to as dynamic instability and dynamic stability, respectively~\cite{tromborg_mode_1997,bogatov_anomalous_1975}.

In order to find the instantaneous modes, the wavenumber $k(\omega,N)=k(\omega_r,N_r)+\Delta k(\omega,N)$ is expanded as, \begin{equation}
\label{eq:wavenumber}    2i\Delta k L\approx \frac{1}{2}(1-i\alpha)\Gamma v_g \left(g(N)-g(N_r)\right)+i(\omega-\omega_r)\tau_L,
\end{equation}
where $\Gamma$ is the confinement factor, $v_g$ the group velocity, $\tau_L=2L/v_g$ is the roundtrip time in the active section, and $\alpha$ is the linewidth enhancement factor. The reference point $(\omega_r,N_r)$ can be any steady-state mode. 

Returning to the instantaneous modes, they can be shown to satisfy 
\begin{equation}
\label{OCcontinue}
\frac{d\omegan}{dN}=\frac{\tau_L}{\tau_L+\tau_R(\omegan)}\times \frac{1}{2}(i+\alpha)\Gamma v_g g_N,
\end{equation}
 where $g_N=g_N(N)$ is the material differential gain. Furthermore, $\tau_R(\omega)$ is a complex time defined the same way as in Ref.~\cite{tronciu_feedback_2021} by,
\begin{eqnarray}\tau_R(\omega) \equiv -i\frac{d}{d\omega}\ln r_R(\omega). \end{eqnarray} Its real part corresponds to an effective roundtrip time in the PDR given by the frequency derivative of the phase. In contrast, the imaginary part leads to additional phase-amplitude coupling. Compared to the case without a PDR, $\frac{d\omega}{dN}$ is modified by a factor $\tau_L/(\tau_L+\tau_R(\omega))$, which appears as a complex-valued weighting or confinement factor. Thus, we can define an effective confinement factor $\tGamma(\omega)$ and an effective linewidth enhancement factor $\talpha(\omega)$ by 
\begin{eqnarray}
\label{eq:eff_parameters}
    \tGamma(\omega)(i+\talpha(\omega))\equiv\frac{\Gamma (i+\alpha)}{1+\tau_R(\omega)/\tau_L}.
\end{eqnarray}
With these definitions, we get, 
\begin{eqnarray}
    \frac{d \omegan}{dN}=\frac{1}{2}(i+\talpha(\omegan))\tGamma(\omegan)v_g g_N.
\end{eqnarray}
Defining a general differential gain function $\oG_N(\omegan,N)\equiv \tGamma(\omegan)v_g g_N(N)$, the modal differential gain for the $n$'th instantaneous mode is $\oG_{Nn}(N)=\oG_N(\omegan_n(N),N)$. Due to the frequency-dependence of the PDR, $\tGamma(\omega)$ and $\talpha(\omega)$ will vary between the different steady-state solutions. The condition for a steady-state solution $(\omega_s,N_s)$ to be an antimode can now be expressed as 
\begin{eqnarray}
        \oG_N(\omega_s,N_s)\propto \tGamma(\omega_s)<0\quad (\text{antimodes}),
\end{eqnarray}
while steady-state modes have $\tGamma(\omega_s)>0$. 

We note that similar effective parameters have been derived in Refs.~\cite{tronciu_feedback_2021,vahala_detuned_1984}, with the slight difference that instead of $\tGamma(\omega)$, Ref.~\cite{vahala_detuned_1984} defines an effective relaxation oscillation frequency, while Ref.~\cite{tronciu_feedback_2021} defines an effective photon lifetime. Here, we take the effective confinement factor to be more fundamental as, arguably, the effect on the photon lifetime and the relaxation oscillation frequency is \textit{because} of the modified confinement in the active section. 
Further, if the PDR is only weakly dispersive, such that $\tau_R(\omega)$ can be approximated by its steady-state value, $\tau_R(\omega_s)$, for a particular steady-state frequency $\omega_s$, the dynamics will be qualitatively similar to a conventional Fabry-Perot laser, but with the rescaled parameters $\tGamma(\omega_s)$ and $\talpha(\omega_s)$.  
In this weakly dispersive limit, the PDR results in a scaling of the modal differential gain $\oG_N=\tGamma(\omega_s)v_g g_N$, the linewidth, the relaxation oscillation frequency $\tomega_R=\sqrt{(R_p/R_{p,th}-1)\oG_NN_s/\tau_s}$, and the photon lifetime $\overline{\tau}_p=\left(\tGamma(\omega_s)v_g g_{th}\right)^{-1}$, where $g_{th}=g(N_{th}(\omega_s))$ is the material threshold gain.
\subsection{Dynamical model}
The model we use is based on the iterative model in Ref.~\cite{detoma_complex_2005} and also used in Ref.~\cite{rimoldi_damping_2022}. We define the slowly varying envelopes $A(t)$ and $A_-(t)$ by 
\begin{eqnarray}
\label{eq:FT}
    A_{(-)}(t) e^{-i\omega_rt}=\frac{1}{2\pi}\int_{0}^{\infty}E_{(-)}(\omega)e^{-i\omega t}d\omega.
\end{eqnarray}

Using this definition of the Fourier transform along with the wavenumber expansion in Eq.~(\ref{eq:wavenumber}), Eq.~(\ref{eq:Eplus}) can be transformed into the time domain to give expressions for $A(t)$ and $A_-(t)$,
\begin{align}
    A(t) =& \, e^{\frac{1}{2}(1-i\alpha)\Gamma v_g \left(\langle g(N)\rangle -g(N_r)\right)\tau_L}\!\times\! \frac{A_-(t-\tau_L)}{r_R(\omega_r)}+F(t)
\end{align}
where $F(t)$ is the inverse Fourier transform of $F(\omega)$, and $\langle g(N)\rangle (t)$ is the gain averaged over one roundtrip in the laser cavity, \begin{eqnarray}
    \langle g(N)\rangle=\frac{1}{\tau_L}\int_{t-\tau_L}^tg(N(t'))dt'.
\end{eqnarray}
The reflected field $A_-(t)$ is given formally by
\begin{eqnarray}
\label{Aminus}
    A_-(t) = \int_{-\infty}^{t}\hat{r}_R(t-t')A(t')dt',
\end{eqnarray}
where $\hat{r}_R(t)$ is the impulse response function of the PDR, which is the inverse Fourier transform of $r_R(\omega)$, using the same definition as in Eq.~(\ref{eq:FT}). We remark that the form of $A_-(t)$ given in Eq.~(\ref{Aminus}) is used solely for analysis. For numerical simulations, it is advantageous to describe $A_-(t)$ in terms of a rate equation derived from, e.g., coupled-mode theory. Finally, the evolution of the carrier density is described by the rate equation \begin{equation}
    \frac{d}{dt}N=R_p-\frac{N}{\tau_s}-v_g g(N)N_p.
\end{equation}
Here, $N_p$ is the photon number density. In steady-state, $N_p\propto |A|^2$ with a proportionality constant given in Ref.~\cite{tromborg_transmission_1987}. Assuming this proportionality to hold out of equilibrium, we normalize $|A|^2=N_p$.

\subsection{Relaxation oscillations} Next, we perform a linear stability analysis of the steady-state modes fulfilling $\tGamma(\omega_s)>0$. A key result regards the impact of a dispersive mirror on the relaxation oscillations. Relaxation oscillations are intensity oscillations that occur due to coupling between the carrier- and photon reservoirs. The linear stability analysis is carried out by assuming perturbations from steady-state with characteristic time-dependence $e^{-i\Omega t}$. The real part $\Re(\Omega)$ gives the angular frequency, and $\Im(\Omega)<0$ corresponds to damped oscillations, while $\Im(\Omega)>0$ means the oscillations are undamped. For a steady-state mode to be stable, no eigenvalue $\Omega$ can have positive imaginary part. 

For conventional Fabry-Perot lasers, the frequency and damping rate of relaxation oscillations are determined by a characteristic equation of the form~\cite{coldren_diode_2012}, \begin{equation}
    -\Omega^2-i\gamma_R\Omega+\omega_R^2=0,
\end{equation}
where the damping rate $\gamma_R$ and relaxation resonance frequency $\omega_R$ are given by
\begin{equation}
  \gamma_R=\frac{1}{\tau_s}+\tau_{p}\omega_R^2, \quad \omega_R^2= \Gamma v_g g_N \left(R_p-\frac{N_s}{\tau_s}\right).
\end{equation}
Here, $\tau_p=1/\Gamma v_g g_{th}$ is the photon lifetime, where $g_{th}=g(N_s)$ is the material threshold gain. 
For lasers with dispersive mirrors, we can derive a generalization of the above characteristic equation to include the effect of a frequency-dependent mirror (see supplementary information for the details)
\begin{equation}   
\label{eq:characteristic}
    -\Omega^2-i\gamma_R\Omega+\frac{1}{2}\omega_R^2\left(H(\Omega)+H^*(-\Omega^*)\right)=0,
\end{equation}
where $H(\Omega)$ is given by\begin{equation}
    H(\Omega) =(1-i\alpha) \frac{e^{i\Omega\tau_L}-1}{\frac{r_R(\omega_s+\Omega)}{r_R(\omega_s)}e^{i\Omega\tau_L}-1}.
\end{equation} 

The combined transfer function $H_P(\Omega)=(H(\Omega)+H^*(-\Omega^*))/2$ relates the power to the carrier density, i.e. $-i\Omega \delta P(\Omega)\propto H_P(\Omega)\delta N(\Omega)$. Physically, the two terms take into account that intensity oscillations at a frequency $\Omega$ generate two optical sidebands at the frequencies $\omega_s\pm\Omega$. If the reflectivity depends on frequency, these optical sidebands will experience different loss and phase delays. 

Equation~(\ref{eq:characteristic}) cannot be solved analytically in general, but if $H(\Omega)$ varies slowly in the vicinity of $\Omega=0$, which is the case for moderately dispersive mirrors, we may define approximate relaxation resonance frequencies and damping rates by 
\begin{align}
\label{effomegaR}    \tomega_R^2&\equiv \omega_R^2\Re(H(0)), \\
\label{effgammaR}
    \tgamma_R &\equiv \gamma_R-\omega_R^2\Im(H'(0)),
\end{align}
 where the prime denotes the derivative with respect to $\Omega$. The characteristic equation then becomes \begin{equation}
    -\Omega^2-i\tgamma_R\Omega+\tomega_R^2=0.
\end{equation} 

Evaluated at $\Omega\rightarrow 0$, we have \begin{equation}
    H(0)=\frac{1-i\alpha}{1+\tau_R(\omega_s)/\tau_L}=\frac{\tGamma(\omega_s)}{\Gamma}\left(1-i\talpha(\omega_s)\right),
\end{equation}
where we recall $\tau_R(\omega)=-i\partial_\omega\ln r_R(\omega)$, and
\begin{eqnarray}
  H'(0)  = \frac{1}{2}\tau_L(\alpha+i)\frac{r_R''(\omega_s)/r_R(\omega_s)}{(\tau_L+\tau_R(\omega_s))^2}.
\end{eqnarray}
Remarkably, the approximation for the relaxation oscillation frequency matches the usual expression if the rescaled confinement factor is used. 

\begin{figure}
    \centering \includegraphics[width=\linewidth]{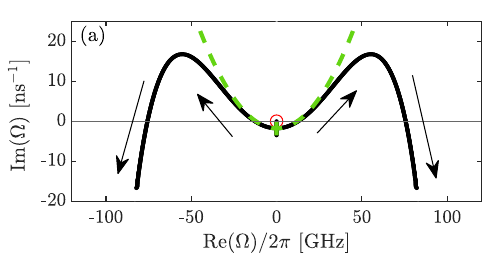} \includegraphics[width=\linewidth]{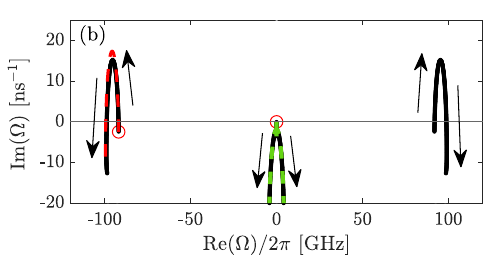}
    \caption{ Evolution of the eigenvalues of the linearization related to the lowest-threshold steady-state mode as the pump rate is varied for the case of (a) dispersive self-Q-switching and (b) beating oscillations. The green dashed curves show the approximation in Eqs.~(\ref{effomegaR}) and~(\ref{effgammaR}), while the red dashed curve shows the approximation in Eq.~(\ref{eq:DeltaOmega}). The red circles show the positions of the sidemodes $\Omega=\omegan_{ms}-\omega_s$.}
    \label{fig:eigenvalues}
\end{figure}

The eigenvalues are in general functions of the pump rate. As an example, in Fig.~\ref{fig:eigenvalues} we consider two sets of parameters for the coupled-cavity Fano laser given in Table~\ref{tab:parameters}, which are also used later in sections \ref{sec:dsq} and \ref{sec:beat}, and plot the traces of the eigenvalues in the complex plane as the pump rate varies. The exact solutions to Eq.~(\ref{eq:characteristic}) are given in black, while the approximations to the relaxation oscillation eigenvalues are given in dashed green, showing good agreement for $\Re(\Omega)/2\pi\lesssim30$ GHz. The points where the eigenvalues cross the real axis, $\Im(\Omega)=0$, correspond to Hopf bifurcations. 
 
This general analytical result can be applied to a large class of lasers and can be used to explain a number of results presented in the literature. Ref.~\cite{rasmussen_modes_2018} considers the case of a single-cavity Fano laser without a blocking hole in the waveguide and where the lasing frequency coincides exactly with the resonance peak of the cavity. Their results on the relaxation oscillation frequency and damping rate agree with the expressions above, and furthermore, we can now easily evaluate the influence of detuning and the presence of a blocking hole. Another example is a laser with weak optical feedback with an effective reflectivity that can be approximated as $r_R(\omega) = r_2(1+\kappa \exp(i\omega\tau))$, where $\kappa$ is the feedback strength and $\tau$ is the roundtrip time in the feedback arm. In this case, \begin{equation}
    \Re(H(0))= \frac{1+\frac{\kappa\tau}{\tau_L}\left(\cos(\omega_s\tau)-\alpha \sin(\omega_s \tau)\right)}{1+\left(\frac{\kappa\tau}{\tau_L}\right)^2+2\frac{\kappa\tau}{\tau_L}\cos(\omega_s\tau)},
\end{equation}
which agrees with the modification of the relaxation oscillation frequency given in Ref.~\cite{agrawal_semiconductor_1993}.

Returning to the general analysis, we observe that the second derivative $r_R''(\omega_s)$ is decisive for the stability. In the case where $|r_R(\omega)|$ is at an extremum, we have $\Im(H'(0))\propto \alpha(\arg r_R)''+(\ln|r_R|)''-(\arg r_R )'^2$. This shows that the damping rate of relaxation oscillations increases near the maximum of a reflection peak while it decreases near a minimum. In fact, the second and third terms have clear physical interpretations, corresponding to spectral filtering/amplification and increased storage time in the passive section. This agrees with the results and physical interpretations given in Ref.~\cite{rasmussen_suppression_2019}, showing that a Fano laser operating at the reflection peak has increased tolerance towards external feedback. 

Finally, a positive curvature of the reflection spectrum leads more readily to instabilities. In Ref.~\cite{ramunno_stability_2002}, the authors analyze self-pulsing in a laser with reflection from a chirped grating, and they also conclude that the phase curvature is decisive for the stability. This insight can be used as a guideline for designing self-pulsing lasers based on dispersive mirrors.

\subsection{Photon-photon resonances}
In addition to the eigenvalues of Eq. (\ref{eq:characteristic}) that relate to relaxation oscillations, another set of eigenvalues is related to coupling between the steady-state mode and sidemodes. We notice that $H(\Omega)$ has poles at the positions of the sidemodes relative to the steady-state mode, $\Omega_{ms}=\omegan_{ms}-\omega_{s}$. Just above threshold, where $\omega_R^2\sim 0^+$ is negligible, the poles $\Omega_{ms}$ are exact eigenvalues. When further increasing the pump rate, the eigenvalues move in the complex plane.

Writing $\Omega= \Omega_{ms}+\Delta \Omega$, we can get the approximate expression
 \begin{equation}
 \label{eq:DeltaOmega}
\Delta \Omega \approx \frac{\frac{1}{2}\omega_{R}^2}{\gamma_{R}-i\Omega_{ms}}\times \frac{1-i\alpha}{1+\tau_R(\omegan_{ms})/\tau_L} \times \frac{e^{i\Omega_{ms}\tau_L}-1}{i\Omega_{ms}\tau_L},
\end{equation}
which is valid for $|\Delta \Omega| \ll |\Omega_{ms}|$. The key point here is that the first factor gives rise to an asymmetric mode-coupling, which dampens sidemodes on the blue side ($\Re(\Omega_{ms})>0$) and amplifies sidemodes on the red side ($\Re(\Omega_{ms})<0$). This four-wave mixing effect, mediated by carrier oscillations, is known as the Bogatov effect~\cite{bogatov_anomalous_1975}. The effect is responsible for the onset of beating oscillations (where the lowest threshold "blue" mode becomes unstable), as well as the termination of beating oscillations due to so-called dynamic stability of the "red" mode with higher threshold~\cite{tromborg_mode_1997}. In Ref.~\cite{marconi_asymmetric_2016}, the authors show that the effect can also be used to transfer energy from the "blue" mode to the "red" mode in the case of two coupled photonic crystal cavities.

Figure~\ref{fig:eigenvalues}b shows an example of a case where a pair of eigenvalues related to photon-photon-resonances cross the real axis, leading to instability. Specifically, we consider the eigenvalues of the lowest threshold steady-state mode for the laser in section~\ref{sec:beat}, which has a weakly damped sidemode near $-93$ GHz (red circle). When the pump rate is increased above the lasing threshold, the eigenvalues related to this sidemode move upwards and become unstable. The approximation given in Eq.~(\ref{eq:DeltaOmega}) is shown as the dashed red line in Fig.~\ref{fig:eigenvalues}b, indicating good agreement with the numerical result. 
 
\section{Application: Self-Q-Switching}
\label{sec:dsq}
\begin{figure}
\includegraphics[width=1\linewidth]{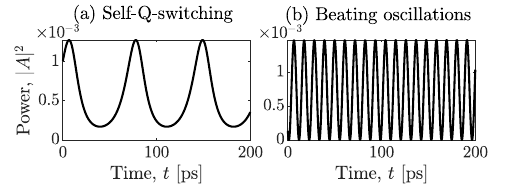}
 \caption{Time-traces of the normalized power in the active section during (a) dispersive self-Q-switching, and (b) beating-type oscillations. In (a), we observe a train of pulses with pulse widths of $\sim$20 ps and repetition rate of $\sim$15 GHz. In (b), we observe beating oscillations, which in the frequency domain corresponds to dual-mode (or two-color) lasing. We observe fast ($\sim $93 GHz) sinusoidal oscillations that result from the beating between two modes lasing simultaneously, which are locked together through carrier oscillations in the active medium.}
     \label{fig:dsqbeat}
\end{figure}

We now apply the general results derived above to the coupled-cavity Fano laser (see Fig.~\ref{fig:1}a), which provides a general platform to tailor the mirror dispersion. First, we consider the possibility of self-Q-switching, i.e., the instability related to relaxation oscillations leading to the self-pulsing shown in Fig.~\ref{fig:dsqbeat}a. 

\begin{table}
    \centering
    \begin{tabular}{lll}
    \textbf{Parameter} & \textbf{Symbol} & \textbf{Value}\\
    \hline
        Material differential gain & $g_N$ & $5\times 10^{-12} $ cm$^2$\\
        \hline
        Transparency carrier density & $N_0$ & $5\times 10^{15}$ cm$^{-3}$ \\
        \hline
        Carrier lifetime & $\tau_s$ & 0.28 ns\\
        \hline
         Confinement factor & $\Gamma$ & 0.01\\
        \hline
        Left mirror reflectivity &  $r_1$ & -0.99\\
        \hline
         Waveguide losses &  $\alpha_i$& 10 cm$^{-1}$\\
        \hline
         Reference length of laser cavity & $L$ & 4.98 $\mu$m\\
        \hline
         Reference refractive index & $n_r$ & 3.5 \\
        \hline
         Reference group index & $n_g$ & 3.5\\
        \hline
         Linewidth enhancement factor & $\alpha$ & 3.3\\
         \hline 
         \textbf{Parameters related to PDR} \\
    \hline
         Reference wavelength & $\lambda_r$ & 1554 nm\\
         \hline
         Cavity 1 resonance frequency &$\omega_{c1}$ & $2\pi c /\lambda_r$\\
         \hline
         Cavity 2 detuning from $\omega_{c1}$ & $\Delta$ & 0 (-100) GHz\\
         \hline
        Vertical scattering $Q$ & $Q_i$ & $10^5$\\
        \hline
        Cavity-waveguide $Q$ & $Q_w$ & 750 \\
        \hline
        Decay rate related to channel $x$ & $\gamma_x$ &$(2\pi c/\lambda_r)/2Q_x$\\
        \hline
        Direct coupling rate & $\mu$ & 0.65 (0) $\gamma_w$\\
        \hline 
        Indirect coupling phase &$\theta$ & -$\pi/6$ (0)\\
    \hline
    Fundamental steady-state mode & $\omega_{s0}$ & $\omega_{c1}$+0.65 $\gamma_w$ ($\omega_{c1}$)\\
    \hline
    
    \end{tabular}
    
    \caption{Parameters used in simulation. Parameters outside (inside) of parentheses are specific to simulations presented in section \ref{sec:dsq} (\ref{sec:beat}). Material parameters are based on Ref.~\cite{mork_semiconductor_2019}.}
    \label{tab:parameters}
\end{table}

\begin{figure*}
    \centering
    \includegraphics[width=0.49\linewidth]{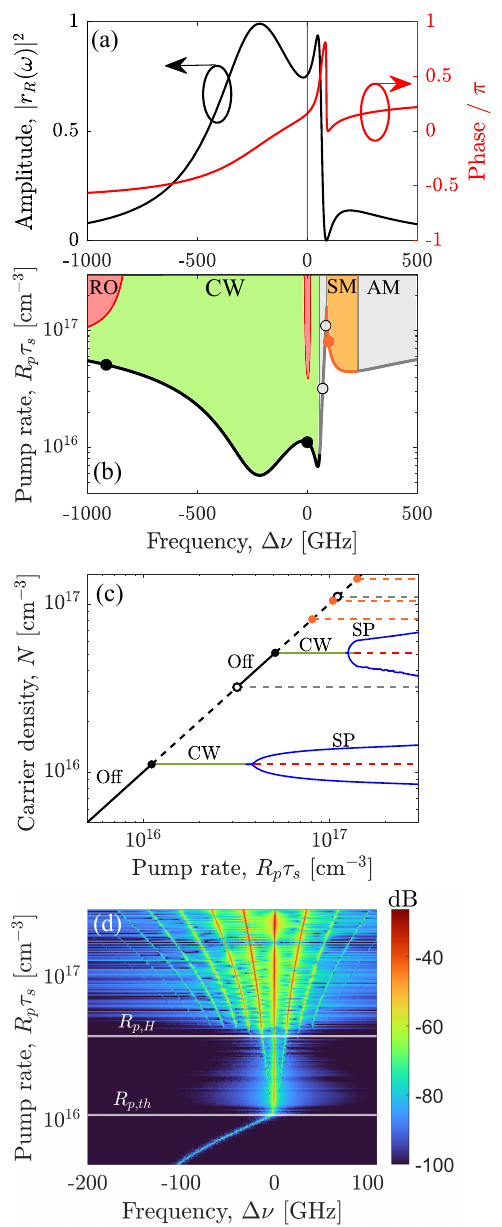}
        \includegraphics[width=0.49\linewidth]{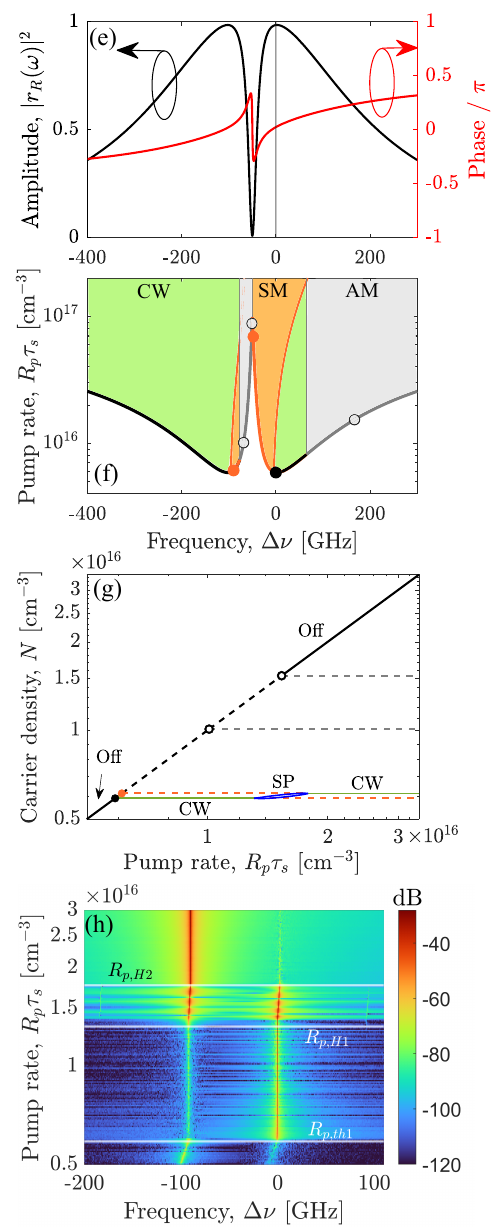}
    \caption{Figures related to self-Q-switching in the first column (a-d) and beating oscillations in the second column (e-h). (a,e) Mirror reflectivities $r_R(\omega)$ with $|r_R(\omega)|^2$ in black (left axes) and $\arg(r_R(\omega))/\pi$ in red (right axes). (b,f) Stability diagram in terms of frequency and pump rate, showing the lasing threshold as a function of frequency, the various steady-state solutions, and the boundary for self-pulsing. Description in the main text. (c,g) Bifurcation diagrams with $N$ vs. $R_p$. (d,h) Optical spectrum in color scale for varying pump rate.}
    \label{fig:stability_dsq}
\end{figure*}

Knowing that we need a positive curvature, we choose parameters that lead to the reflection spectrum in Fig.~\ref{fig:stability_dsq}a, where $|r_R(\omega)|^2$ is shown in black (left axis) and $\phi(\omega)$ is shown in red (right axis). 
The frequency axis is $\Delta \nu = (\omega-\omega_{s_0})/2\pi$, where the vertical line at $\Delta\nu=0$ indicates the targeted steady-state frequency $\omega_{s0}$ used in simulations unless otherwise specified.  

Figure~\ref{fig:stability_dsq}b shows a stability diagram of steady-state solutions in frequency $\omega_s$ and pump rate $R_p$. The white region is below the lasing threshold $R_p<N_{th}(\omega)/\tau_s$, while the colored regions are above threshold. Further, the color indicates the stability of steady-state solutions in that region, and if unstable, whether the undamped eigenvalues ($\Im(\Omega)>0$) can be attributed to relaxation oscillations (red, RO), unsuppressed sidemodes (orange, SM), or antimodes (AM, grey). 
Green (CW) denotes regions of stable CW lasing where all eigenvalues have negative imaginary parts. We indeed observe that the self-pulsing threshold for undamped relaxation oscillations is lowest near the point where $|r_R(\omega)|^2$ has the largest positive curvature. 
Additionally, the circle markers on the lasing threshold border indicate the frequencies $\{\omega_s\}$ of the particular set of steady-state solutions that coexist with a steady-state mode at the indicated target frequency $\omega_{s} =\omega_{s0}$. That is, inserting $\omega_s=\omega_{s0}$ in the phase condition Eq.~(\ref{eq:phasecondition}) fixes the required global phase $\phi_0$, which in turn determines all other steady-state solutions. 

Figure~\ref{fig:stability_dsq}c shows the bifurcation diagram in the ($R_p$,$N$)-plane for a choice of global phase giving rise to the steady-state solutions indicated in Fig.~\ref{fig:stability_dsq}b. As the pump rate varies, the lines show the position and stability of the various fixed points and limit cycles. A fully drawn line denotes a stable state with the type being indicated (Off-state in black, CW in green, and self-pulsing in blue). A dashed line represents an unstable state. The black diagonal, $N=R_p\tau_s$, is the Off-state corresponding to a solution with zero photons in the cavity. It is stable if all the instantaneous modes are below threshold at $N=R_p\tau_s$, i.e. $\Im(\omegan_n(R_p\tau_s))<0$ for all $n$. The horizontal lines are steady-state solutions, with steady-state modes in black and antimodes in gray. For the self-pulsing states, the blue curves denote the maximum and minimum carrier density of the limit cycle. Note that we are only considering deterministic dynamics here, so spontaneous emission noise is neglected. Finally, we note that the curve corresponding to the limit cycle at higher carrier density is slightly jagged due to the dynamics not being a simple period-one limit cycle but containing additional frequency components.  

We observe that the lowest threshold mode is initially globally attracting but becomes unstable through a Hopf-bifurcation, giving rise to self-pulsing. Additionally, we observe the existence of pump regimes where 1) a stable CW-state coexists with a stable OFF-state, 2) a stable limit cycle coexists with a stable OFF-state, 3) a stable limit cycle coexists with a stable CW-state, and 4) only non-stationary states exist. This shows the rich landscape of possible dynamics in the system. The various kinds of multistable regimes could also be attractive for switching applications or for excitable behaviour~\cite{wunsche_excitability_2001}.

Figure~\ref{fig:stability_dsq}d shows a 2D map of the evolution of the optical spectrum as a function of the pump rate, computed by numerical integration of the dynamical equations and including Langevin noise. Below the threshold for self-pulsing, the two sidebands arising from relaxation oscillations can be observed. Above the self-pulsing threshold at $R_p\approx 3.5R_{p,th}$, we observe the formation of a frequency comb. As the pump rate further increases, more and more comb lines become visible, which can be interpreted as four-wave mixing of the relaxation oscillation sidebands and the carrier frequency. In the time domain, this corresponds to self-pulsing, which manifests as sinusoidal temporal modulation of the power at the onset of the instability, evolving into a train of pulses when the pump rate is increased. The dynamics are shown in Fig.~\ref{fig:dsqbeat}a, for a pump rate well above the onset of self-pulsing. Typical pulse widths are on the order of $\sim$10-20 ps, and repetition rates are on the order of $\sim$ 10-20 GHz.

\section{Application: Beating Oscillations}
\label{sec:beat}
The second application we consider is the realization of a dual-mode laser, which exhibits beating oscillations (see Fig.~\ref{fig:dsqbeat}b). This type of self-pulsing consists of fast sinusoidal oscillations with a frequency in the range of $\sim 20-200$ GHz, which can be tuned by the design of the passive reflector. Physically, this type of dynamics is interpreted as the beating between two modes that lase simultaneously and are locked together through carrier oscillations in the active medium. The requirement for beating oscillations is a weakly damped sidemode, which is red-detuned from the lowest threshold steady-state mode~\cite{bogatov_anomalous_1975,tromborg_mode_1997}.

We achieve beating oscillations with the reflectivity spectrum in Fig.~\ref{fig:stability_dsq}e, where the two cavity supermodes are spectrally aligned, thus inducing a narrow transparency window where they overlap due to destructive interference. As such, this is an example of electromagnetically induced transparency (EIT) in optical microcavities~\cite{liu_electromagnetically_2017}. The EIT resonance arises due to strong Fano interference, that is, interference between different optical pathways. 
Importantly, the EIT resonance is accompanied by a small wiggle or undulation of the phase response. Due to the negative group delay within the transparency window, the phase condition can be satisfied on both sides of the transparency window. 

In Fig.~\ref{fig:stability_dsq}f, we observe two steady-state modes with comparable thresholds, with the mode on the blue side (black dot) having a slightly lower threshold. 
We observe that the steady-state mode at $\Delta \nu=0$ becomes unstable above a certain threshold when the pump rate crosses into the orange region. On the other hand, the steady-state mode in the orange region on the red side, which is initially unstable, becomes stable at even higher pump rates. Of course, this boundary collides with the laser threshold curve at the point where the two modes have identical thresholds. 

Conversely, when the lowest threshold mode is on the red side, it is always stable. In the present case, the coupling between these two modes is what gives rise to the dual-mode operation. Further, the beat note frequency is approximately given by the detuning $\Delta$ between the two cavities constituting the dispersive laser mirror. This parameter can be tuned dynamically by electro-optic means or through design by modifying the cavity geometry. 

Figure~\ref{fig:stability_dsq}g shows the bifurcation diagram. Again, we observe various kinds of multistable behavior, but most importantly, there is a window between the instability of the lowest threshold mode and the dynamic stability of the second lowest mode where beating oscillations occur. 

Figure~\ref{fig:stability_dsq}h shows the optical spectrum. The "blue" mode starts to lase at $R_{p,th1}$, and we observe single-mode lasing until $R_p=R_{p,H1}$, which constitutes the onset of dual-mode lasing. At $R_p=R_{p,H2}$, the "red" mode becomes stable, and the dual-mode lasing stops. Due to the inclusion of Langevin noise in the simulations, outlines of both the "red" and "blue" modes can be observed even when they don't lase.

Finally, we remark that the two steady-state modes are resonant with different cavities, meaning that the photon densities in the two cavities differ strongly and depend on the oscillating mode. Therefore, if the two cavities are coupled to different cross-ports, the two steady-state modes will have different output channels. Deliberately inducing mode-hopping could then be used as a routing mechanism.
\section{Discussion} 
Comparing the coupled-cavity Fano laser to earlier demonstrations and predictions of self-pulsing lasers, we wish to highlight a few points. In Ref.~\cite{hamel_spontaneous_2015}, the authors demonstrate a symmetric coupled-cavity laser system consisting of two coupled photonic crystal lasers, which exhibits spontaneous mirror symmetry breaking above a certain pump rate, where the energy distribution will become asymmetric and mainly concentrated in a single cavity. Additionally, they predict regimes where the power will spontaneously oscillate back and forth between the two nanocavities. Compared to the present case, a major difference is that the two nanocavities are both active and form two "complete" laser cavities, while in our case, the coupled cavities are passive and merely work as a frequency-dependent mirror. The mechanism for self-pulsing in Ref.~\cite{hamel_spontaneous_2015} is thus attributed to an AC Josephson-like effect and not undamped relaxation oscillations or mode-beatings as in our case.  

If we compare the coupled-cavity Fano laser in the self-Q-switched regime to other self-pulsing lasers with dispersive mirrors, such as the hybrid laser in Ref.~\cite{rimoldi_damping_2022} or qualitatively similar devices in Refs.~\cite{mak_linewidth_2019,huang_high-power_2019}, the frequency combs also emerge through undamped relaxation oscillations. Due to the size of those macroscopic structures, they also produce much higher output power. On the other hand, the Fano laser allows extreme miniaturization and a lower threshold. It should be noted that in contrast to the dispersive self-Q-switching discussed in section~\ref{sec:dsq}, the beating-type oscillations can also occur in lasers that are not based on PDRs. An example is the self-pulsing square-microcavity laser in Ref.~\cite{li_self-pulsing_2023}, where mode-coupling occurs through spatial hole-burning, leading to a spatial modulation of the refractive index. 
\section{Conclusion}
In conclusion, we have presented a general analysis of lasers incorporating a passive dispersive mirror. The analysis can be used to study a large class of lasers, and the insights can be applied to tailor the dynamics of lasers. In particular, we applied the model to a new laser geometry, which provides a flexible platform for realizing many different types of cavity dispersion. 
In combination with simulations, e.g. using FDTD, of the reflection spectrum in specific devices, the model can be used as a design guideline. Finally, the model may serve as a starting point for extending the formalism to study \textit{active} dispersive mirrors, where the gain and refractive index vary in time due to nonlinearities in the cavities or the presence of active material.
\begin{acknowledgments}
This work was supported by the Danish National Research Foundation (Grant No. DNRF147 - NanoPhoton) and by the European Research Council (Grant No. 834410 FANO). 
M.H. and Y.Y. acknowledge the support from Villum Foundation via the Young Investigator Programme (Grant No. 37417 - QNET-NODES, and Grant No. 42026 - EXTREME).
\end{acknowledgments}
\appendix

\section{The coupled-cavity Fano mirror}

In this section we consider a system of a waveguide with two side-coupled cavities as in Fig. 1 in the main text, and apply temporal coupled-mode theory \cite{fan_temporal_2003,wonjoo_suh_temporal_2004} to calculate the reflection spectrum of the effective mirror. The temporal coupled-mode equations describe the slowly-varying envelopes $a_1(t)$ and $a_2(t)$ of the modes of the individual cavities.

The governing equations are, \begin{align}
\label{eq:1}\frac{d}{dt}\avec(t)&= -i\left(\Omvec -\uun{I}\omega_r\right)\avec(t)+\dvec A_+(t),\\
A_-(t) &= \dvec^T \avec(t),
\end{align}
where $\avec = (a_1,a_2)^T$, $A_+(t)$ and $A_-(t)$ are the incoming- and outgoing fields, $\omega_r$ is a reference frequency, $\dvec = (\sqrt{\gamma_{c1}}e^{i\theta_{c1}},\sqrt{\gamma_{c2}}e^{i\theta_{c2}})^T$ is the vector of coupling constants between waveguide and cavities, and $\Omvec$ is the system matrix given by
\begin{align}
    \Omvec &= \begin{pmatrix} \omega_{c1}-i\gamma_{c1}-i\gamma_{i1}&\mu+i\sqrt{\gamma_{c1}\gamma_{c2}}e^{i\theta} \\
    \mu+i\sqrt{\gamma_{c1}\gamma_{c2}}e^{i\theta} & \omega_{c2}-i\gamma_{c2}-i\gamma_{i2}\end{pmatrix}. 
\end{align}
Here $\omega_{c1,2}$ are resonance frequencies of the bare cavities, $\gamma_{c1,2}$ are decay rates related to coupling to the waveguide, $\gamma_{i1,2}$ are decay rates related to intrinsic losses (e.g., vertical scattering), $\mu$ is a direct coupling rate assumed real. 

Transforming (\ref{eq:1}) into the frequency domain, using the definition $e^{-i(\omega-\omega_r)t}$, gives the following solution, \begin{equation}
    \avec(\omega)=\left(i(\Omvec-\uun{I}\omega)\right)^{-1}\dvec A_+(\omega),
\end{equation}
and the effective reflectivity $r_R(\omega)$ becomes
\begin{equation}
    r_R(\omega)=\dvec^T\left(i(\Omvec-\uun{I}\omega)\right)^{-1}\dvec.
\end{equation}
As long as the eigenvalues of $\Omvec$ are non-degenerate, $r_R(\omega)$ can be expanded as \begin{equation}
    r_R(\omega) = \frac{d_+}{i(\omega_+-\omega)+\gamma_+}+\frac{d_-}{i(\omega_--\omega)+\gamma_-},
\end{equation}
where $\omega_\pm-i\gamma_\pm$ are the eigenvalues of $\Omvec$. 
\section{Stability analysis}
In this section, we derive the characteristic equation (\ref{eq:characteristic}), starting from equations (16-20) in the main text. For simplicity, we assume that the contribution of spontaneous emission into the lasing mode is negligible, such that the exact CW solutions correspond to the solutions of the oscillation condition $r_R(\omega_s)r_L(\omega_s,N_s)=1$. The Langevin-noise term $F(t)$ can then be treated as a driving term in the linearized equations. The CW solutions are then given by \begin{align}
    A(t)&=A_se^{-i(\omega_s-\omega_r)t},\\
    A_-(t)&=r_R(\omega_s)A_se^{-i(\omega_s-\omega_r)t}, \\
    N(t) &= N_s. 
\end{align}
Setting $\frac{d}{dt}N(t)=0$ gives a relation between $N_s$ and $A_s$, 
\begin{equation}
    R_p -\frac{N_s}{\tau_s}-v_g g(N_s)|A_s|^2 = 0,
\end{equation}
where without loss of generality, $A_s$ can be taken as real. 

We now linearize the system around a particular steady-state solution,
\begin{align}
    A(t) &= \left(A_s+\delta A(t)\right)e^{-i(\omega_s-\omega_r)t}, \\
    A_-(t) &= \left(r_R(\omega_s)A_s+\delta A_-(t)\right)e^{-i(\omega_s-\omega_r)t},\\
    N(t) &= N_s + \delta N(t).
\end{align}

Inserting in the dynamical equations (16-20), and keeping only terms linear in $\delta$, we get the linearized system   
\begin{eqnarray}
    \delta A(t) &=&\frac{1}{2}(1-i\alpha)\Gamma v_g g_N A_s\tau_L \langle \delta N(t)\rangle A_s\nonumber\\*
    &&+\, \frac{\delta A_-(t-\tau_L)}{r_R(\omega_s)} +F(t),\\
    \frac{d}{dt}\delta N(t)&=&-\left(\frac{1}{\tau_s}+v_g g_N|A_s|^2\right)\delta N(t)\nonumber\\*
    &&-v_g g_{th}A_s\left( \delta A(t)+\delta A^*(t)\right)+ F_N(t),
\end{eqnarray}
where the brackets in $\langle \delta N(t)\rangle$ denotes the same time-average as in eq. (17) in the main text. We also added a potential driving term $F_N(t)$ in the carrier density equation. 

In the complex frequency domain, the equations become \begin{eqnarray}
\label{deltaA}
    -X(\Omega)\delta A(\Omega)&=&(1-i\alpha)\gamma_{AN} O(\Omega) \delta N(\Omega)+ F(\Omega),\\
    -i\Omega\delta N(\Omega)&=& -\gamma_R \delta N(\Omega)-\frac{1}{2}\gamma_{NA}\left(\delta A(\Omega)+\delta A^*(-\Omega^*)\right)\nonumber\\*
    &&\,+F_N(\Omega),
\end{eqnarray}
where $\gamma_R = \frac{1}{\tau_s}+v_g g_N|A_s|^2$, $\gamma_{AN} =\frac{1}{2}\Gamma v_g g_N A_s$, $\gamma_{NA} =2v_g g_{th}A_s$, and \begin{equation}
    X(\Omega)=\frac{r_R(\omega_s+\Omega)}{r_R(\omega_s)}e^{i\Omega \tau_L}-1,\quad O(\Omega) = \frac{e^{i\Omega \tau_L}-1}{i\Omega}. 
\end{equation}
A third equation for $\delta A^*(-\Omega^*)$ is obtained by replacing $\Omega$ by $-\Omega^*$ in eq. (\ref{deltaA}) and then taking the complex conjugate. Letting an overline denote the combined operation $\overline{f(\Omega)}=f^*(-\Omega^*)$, and then omitting the arguments, the equations are put on matrix form
\begin{equation}
    \begin{pmatrix}
        -X&0 &\gamma_{AN}(1-i\alpha)O\\
        0 & -\overline{X}& \gamma_{AN}(1+i\alpha)O\\
        \frac{1}{2}\gamma_{NA}&\frac{1}{2}\gamma_{NA} & -i\Omega+\gamma_R
    \end{pmatrix}\begin{pmatrix}
        \delta A\\
        \overline{\delta A}\\
        \delta N
    \end{pmatrix}=\begin{pmatrix}
        F\\
        \overline{F}\\
        F_N
    \end{pmatrix}.
\end{equation}
Taking the determinant of the matrix yields \begin{equation}
\label{eq:det}
    D(\Omega)=X \overline{X}(-i\Omega +\gamma_R)+\frac{1}{2}\omega_R^2O\left(X(1+i\alpha)+\overline{X}(1-i\alpha)\right),
\end{equation}
where $\omega_R^2 = \gamma_{AN}\gamma_{NA}$. The zeros of the system determinant give the eigenvalues. The real part gives the oscillation frequency of the perturbation, and the imaginary part gives the growth rate, which is negative if the perturbation is damped. Note that since $X(0)=X^*(0)=0$, then $\Omega=0$ is always an eigenvalue. The fact that it is \textit{always} an eigenvalue represents the fact that the system is invariant with respect to a global phase.

Finally, for $X,\overline{X}\neq 0$, which is the typical case above threshold $R_p>N_s/\tau_s$, we multiply the determinant with $-i\Omega/X\overline{X}$ and arrive at the characteristic equation (22) in the main text, 
\begin{align}
    -\Omega^2 -i\gamma_R \Omega+\frac{1}{2}\omega_R^2\left(H+\overline{H}\right)&=0,\\
    H(\Omega) =  (1-i\alpha) \frac{e^{i\Omega\tau_L}-1}{\frac{r_R(\omega_s+\Omega)}{r_R(\omega_s)}e^{i\Omega\tau_L}-1}.
\end{align}
The characteristic equation is now cast in a form that resembles the usual case for a Fabry-Perot laser \cite{coldren_diode_2012}.

\bibliography{references}

\begin{thebibliography}{49}%
\makeatletter
\providecommand \@ifxundefined [1]{%
 \@ifx{#1\undefined}
}%
\providecommand \@ifnum [1]{%
 \ifnum #1\expandafter \@firstoftwo
 \else \expandafter \@secondoftwo
 \fi
}%
\providecommand \@ifx [1]{%
 \ifx #1\expandafter \@firstoftwo
 \else \expandafter \@secondoftwo
 \fi
}%
\providecommand \natexlab [1]{#1}%
\providecommand \enquote  [1]{``#1''}%
\providecommand \bibnamefont  [1]{#1}%
\providecommand \bibfnamefont [1]{#1}%
\providecommand \citenamefont [1]{#1}%
\providecommand \href@noop [0]{\@secondoftwo}%
\providecommand \href [0]{\begingroup \@sanitize@url \@href}%
\providecommand \@href[1]{\@@startlink{#1}\@@href}%
\providecommand \@@href[1]{\endgroup#1\@@endlink}%
\providecommand \@sanitize@url [0]{\catcode `\\12\catcode `\$12\catcode `\&12\catcode `\#12\catcode `\^12\catcode `\_12\catcode `\%12\relax}%
\providecommand \@@startlink[1]{}%
\providecommand \@@endlink[0]{}%
\providecommand \url  [0]{\begingroup\@sanitize@url \@url }%
\providecommand \@url [1]{\endgroup\@href {#1}{\urlprefix }}%
\providecommand \urlprefix  [0]{URL }%
\providecommand \Eprint [0]{\href }%
\providecommand \doibase [0]{https://doi.org/}%
\providecommand \selectlanguage [0]{\@gobble}%
\providecommand \bibinfo  [0]{\@secondoftwo}%
\providecommand \bibfield  [0]{\@secondoftwo}%
\providecommand \translation [1]{[#1]}%
\providecommand \BibitemOpen [0]{}%
\providecommand \bibitemStop [0]{}%
\providecommand \bibitemNoStop [0]{.\EOS\space}%
\providecommand \EOS [0]{\spacefactor3000\relax}%
\providecommand \BibitemShut  [1]{\csname bibitem#1\endcsname}%
\let\auto@bib@innerbib\@empty
\bibitem [{\citenamefont {Hu}\ and\ \citenamefont {Oxenløwe}(2021)}]{hu_chip-based_2021}%
  \BibitemOpen
  \bibfield  {author} {\bibinfo {author} {\bibfnamefont {H.}~\bibnamefont {Hu}}\ and\ \bibinfo {author} {\bibfnamefont {L.~K.}\ \bibnamefont {Oxenløwe}},\ }\bibfield  {title} {{\selectlanguage {en}\bibinfo {title} {Chip-based optical frequency combs for high-capacity optical communications}},\ }\href {https://doi.org/10.1515/nanoph-2020-0561} {\bibfield  {journal} {\bibinfo  {journal} {Nanophotonics}\ }\textbf {\bibinfo {volume} {10}},\ \bibinfo {pages} {1367} (\bibinfo {year} {2021})}\BibitemShut {NoStop}%
\bibitem [{\citenamefont {Udem}\ \emph {et~al.}(2002)\citenamefont {Udem}, \citenamefont {Holzwarth},\ and\ \citenamefont {Hänsch}}]{udem_optical_2002}%
  \BibitemOpen
  \bibfield  {author} {\bibinfo {author} {\bibfnamefont {T.}~\bibnamefont {Udem}}, \bibinfo {author} {\bibfnamefont {R.}~\bibnamefont {Holzwarth}},\ and\ \bibinfo {author} {\bibfnamefont {T.~W.}\ \bibnamefont {Hänsch}},\ }\bibfield  {title} {{\selectlanguage {en}\bibinfo {title} {Optical frequency metrology}},\ }\href {https://doi.org/10.1038/416233a} {\bibfield  {journal} {\bibinfo  {journal} {Nature}\ }\textbf {\bibinfo {volume} {416}},\ \bibinfo {pages} {233} (\bibinfo {year} {2002})}\BibitemShut {NoStop}%
\bibitem [{\citenamefont {Feiste}\ \emph {et~al.}(1994)\citenamefont {Feiste}, \citenamefont {As},\ and\ \citenamefont {Ehrhardt}}]{feiste_18_1994}%
  \BibitemOpen
  \bibfield  {author} {\bibinfo {author} {\bibfnamefont {U.}~\bibnamefont {Feiste}}, \bibinfo {author} {\bibfnamefont {D.}~\bibnamefont {As}},\ and\ \bibinfo {author} {\bibfnamefont {A.}~\bibnamefont {Ehrhardt}},\ }\bibfield  {title} {\bibinfo {title} {18 {GHz} all-optical frequency locking and clock recovery using a self-pulsating two-section {DFB}-laser},\ }\href {https://doi.org/10.1109/68.265905} {\bibfield  {journal} {\bibinfo  {journal} {IEEE Photonics Technology Letters}\ }\textbf {\bibinfo {volume} {6}},\ \bibinfo {pages} {106} (\bibinfo {year} {1994})}\BibitemShut {NoStop}%
\bibitem [{\citenamefont {Suh}\ \emph {et~al.}(2016)\citenamefont {Suh}, \citenamefont {Yang}, \citenamefont {Yang}, \citenamefont {Yi},\ and\ \citenamefont {Vahala}}]{suh_microresonator_2016}%
  \BibitemOpen
  \bibfield  {author} {\bibinfo {author} {\bibfnamefont {M.-G.}\ \bibnamefont {Suh}}, \bibinfo {author} {\bibfnamefont {Q.-F.}\ \bibnamefont {Yang}}, \bibinfo {author} {\bibfnamefont {K.~Y.}\ \bibnamefont {Yang}}, \bibinfo {author} {\bibfnamefont {X.}~\bibnamefont {Yi}},\ and\ \bibinfo {author} {\bibfnamefont {K.~J.}\ \bibnamefont {Vahala}},\ }\bibfield  {title} {{\selectlanguage {en}\bibinfo {title} {Microresonator soliton dual-comb spectroscopy}},\ }\href {https://doi.org/10.1126/science.aah6516} {\bibfield  {journal} {\bibinfo  {journal} {Science}\ }\textbf {\bibinfo {volume} {354}},\ \bibinfo {pages} {600} (\bibinfo {year} {2016})}\BibitemShut {NoStop}%
\bibitem [{\citenamefont {Pammi}\ and\ \citenamefont {Barbay}(2020)}]{pammi_micro-lasers_2020}%
  \BibitemOpen
  \bibfield  {author} {\bibinfo {author} {\bibfnamefont {V.~A.}\ \bibnamefont {Pammi}}\ and\ \bibinfo {author} {\bibfnamefont {S.}~\bibnamefont {Barbay}},\ }\bibfield  {title} {\bibinfo {title} {Micro-lasers for neuromorphic computing},\ }\href {https://doi.org/10.1051/photon/202010426} {\bibfield  {journal} {\bibinfo  {journal} {Photoniques}\ ,\ \bibinfo {pages} {26}} (\bibinfo {year} {2020})}\BibitemShut {NoStop}%
\bibitem [{\citenamefont {Shastri}\ \emph {et~al.}(2021)\citenamefont {Shastri}, \citenamefont {Tait}, \citenamefont {Ferreira~de Lima}, \citenamefont {Pernice}, \citenamefont {Bhaskaran}, \citenamefont {Wright},\ and\ \citenamefont {Prucnal}}]{shastri_photonics_2021}%
  \BibitemOpen
  \bibfield  {author} {\bibinfo {author} {\bibfnamefont {B.~J.}\ \bibnamefont {Shastri}}, \bibinfo {author} {\bibfnamefont {A.~N.}\ \bibnamefont {Tait}}, \bibinfo {author} {\bibfnamefont {T.}~\bibnamefont {Ferreira~de Lima}}, \bibinfo {author} {\bibfnamefont {W.~H.~P.}\ \bibnamefont {Pernice}}, \bibinfo {author} {\bibfnamefont {H.}~\bibnamefont {Bhaskaran}}, \bibinfo {author} {\bibfnamefont {C.~D.}\ \bibnamefont {Wright}},\ and\ \bibinfo {author} {\bibfnamefont {P.~R.}\ \bibnamefont {Prucnal}},\ }\bibfield  {title} {{\selectlanguage {en}\bibinfo {title} {Photonics for artificial intelligence and neuromorphic computing}},\ }\href {https://doi.org/10.1038/s41566-020-00754-y} {\bibfield  {journal} {\bibinfo  {journal} {Nature Photonics}\ }\textbf {\bibinfo {volume} {15}},\ \bibinfo {pages} {102} (\bibinfo {year} {2021})}\BibitemShut {NoStop}%
\bibitem [{\citenamefont {Miller}(2017)}]{miller_attojoule_2017}%
  \BibitemOpen
  \bibfield  {author} {\bibinfo {author} {\bibfnamefont {D.~A.~B.}\ \bibnamefont {Miller}},\ }\bibfield  {title} {\bibinfo {title} {Attojoule {Optoelectronics} for {Low}-{Energy} {Information} {Processing} and {Communications}},\ }\href {https://doi.org/10.1109/JLT.2017.2647779} {\bibfield  {journal} {\bibinfo  {journal} {Journal of Lightwave Technology}\ }\textbf {\bibinfo {volume} {35}},\ \bibinfo {pages} {346} (\bibinfo {year} {2017})}\BibitemShut {NoStop}%
\bibitem [{\citenamefont {Yu}\ \emph {et~al.}(2017)\citenamefont {Yu}, \citenamefont {Xue}, \citenamefont {Semenova}, \citenamefont {Yvind},\ and\ \citenamefont {Mork}}]{yu_demonstration_2017}%
  \BibitemOpen
  \bibfield  {author} {\bibinfo {author} {\bibfnamefont {Y.}~\bibnamefont {Yu}}, \bibinfo {author} {\bibfnamefont {W.}~\bibnamefont {Xue}}, \bibinfo {author} {\bibfnamefont {E.}~\bibnamefont {Semenova}}, \bibinfo {author} {\bibfnamefont {K.}~\bibnamefont {Yvind}},\ and\ \bibinfo {author} {\bibfnamefont {J.}~\bibnamefont {Mork}},\ }\bibfield  {title} {{\selectlanguage {en}\bibinfo {title} {Demonstration of a self-pulsing photonic crystal {Fano} laser}},\ }\href {https://doi.org/10.1038/nphoton.2016.248} {\bibfield  {journal} {\bibinfo  {journal} {Nature Photonics}\ }\textbf {\bibinfo {volume} {11}},\ \bibinfo {pages} {81} (\bibinfo {year} {2017})}\BibitemShut {NoStop}%
\bibitem [{\citenamefont {Delmulle}\ \emph {et~al.}(2022)\citenamefont {Delmulle}, \citenamefont {Bazin}, \citenamefont {Massaro}, \citenamefont {Sagnes}, \citenamefont {Pantzas}, \citenamefont {Combrié}, \citenamefont {Raineri},\ and\ \citenamefont {De~Rossi}}]{delmulle_self-pulsing_2022}%
  \BibitemOpen
  \bibfield  {author} {\bibinfo {author} {\bibfnamefont {M.}~\bibnamefont {Delmulle}}, \bibinfo {author} {\bibfnamefont {A.}~\bibnamefont {Bazin}}, \bibinfo {author} {\bibfnamefont {L.~M.}\ \bibnamefont {Massaro}}, \bibinfo {author} {\bibfnamefont {I.}~\bibnamefont {Sagnes}}, \bibinfo {author} {\bibfnamefont {K.}~\bibnamefont {Pantzas}}, \bibinfo {author} {\bibfnamefont {S.}~\bibnamefont {Combrié}}, \bibinfo {author} {\bibfnamefont {F.}~\bibnamefont {Raineri}},\ and\ \bibinfo {author} {\bibfnamefont {A.}~\bibnamefont {De~Rossi}},\ }\bibfield  {title} {{\selectlanguage {en}\bibinfo {title} {Self-{Pulsing} {Nanobeam} {Photonic} {Crystal} {Laser}}},\ }in\ \href {https://doi.org/10.1364/CLEO_AT.2022.JW3B.23} {{\selectlanguage {en}\emph {\bibinfo {booktitle} {Conference on {Lasers} and {Electro}-{Optics}}}}}\ (\bibinfo  {publisher} {Optica Publishing Group},\ \bibinfo {address} {San Jose, California},\ \bibinfo {year} {2022})\ p.\ \bibinfo {pages} {JW3B.23}\BibitemShut {NoStop}%
\bibitem [{\citenamefont {Tronciu}\ \emph {et~al.}(2003)\citenamefont {Tronciu}, \citenamefont {Yamada}, \citenamefont {Ohno}, \citenamefont {Ito}, \citenamefont {Kawakami},\ and\ \citenamefont {Taneya}}]{tronciu_self-pulsation_2003}%
  \BibitemOpen
  \bibfield  {author} {\bibinfo {author} {\bibfnamefont {V.}~\bibnamefont {Tronciu}}, \bibinfo {author} {\bibfnamefont {M.}~\bibnamefont {Yamada}}, \bibinfo {author} {\bibfnamefont {T.}~\bibnamefont {Ohno}}, \bibinfo {author} {\bibfnamefont {S.}~\bibnamefont {Ito}}, \bibinfo {author} {\bibfnamefont {T.}~\bibnamefont {Kawakami}},\ and\ \bibinfo {author} {\bibfnamefont {M.}~\bibnamefont {Taneya}},\ }\bibfield  {title} {{\selectlanguage {en}\bibinfo {title} {Self-pulsation in an {InGaN} laser-theory and experiment}},\ }\href {https://doi.org/10.1109/JQE.2003.819541} {\bibfield  {journal} {\bibinfo  {journal} {IEEE Journal of Quantum Electronics}\ }\textbf {\bibinfo {volume} {39}},\ \bibinfo {pages} {1509} (\bibinfo {year} {2003})}\BibitemShut {NoStop}%
\bibitem [{\citenamefont {Wenzel}\ \emph {et~al.}(1996)\citenamefont {Wenzel}, \citenamefont {Bandelow}, \citenamefont {Wunsche},\ and\ \citenamefont {Rehberg}}]{wenzel_mechanisms_1996}%
  \BibitemOpen
  \bibfield  {author} {\bibinfo {author} {\bibfnamefont {H.}~\bibnamefont {Wenzel}}, \bibinfo {author} {\bibfnamefont {U.}~\bibnamefont {Bandelow}}, \bibinfo {author} {\bibfnamefont {H.-J.}\ \bibnamefont {Wunsche}},\ and\ \bibinfo {author} {\bibfnamefont {J.}~\bibnamefont {Rehberg}},\ }\bibfield  {title} {\bibinfo {title} {Mechanisms of fast self pulsations in two-section {DFB} lasers},\ }\href {https://doi.org/10.1109/3.481922} {\bibfield  {journal} {\bibinfo  {journal} {IEEE Journal of Quantum Electronics}\ }\textbf {\bibinfo {volume} {32}},\ \bibinfo {pages} {69} (\bibinfo {year} {1996})}\BibitemShut {NoStop}%
\bibitem [{\citenamefont {Rimoldi}\ \emph {et~al.}(2022)\citenamefont {Rimoldi}, \citenamefont {Columbo}, \citenamefont {Bovington}, \citenamefont {Romero-García},\ and\ \citenamefont {Gioannini}}]{rimoldi_damping_2022}%
  \BibitemOpen
  \bibfield  {author} {\bibinfo {author} {\bibfnamefont {C.}~\bibnamefont {Rimoldi}}, \bibinfo {author} {\bibfnamefont {L.~L.}\ \bibnamefont {Columbo}}, \bibinfo {author} {\bibfnamefont {J.}~\bibnamefont {Bovington}}, \bibinfo {author} {\bibfnamefont {S.}~\bibnamefont {Romero-García}},\ and\ \bibinfo {author} {\bibfnamefont {M.}~\bibnamefont {Gioannini}},\ }\bibfield  {title} {{\selectlanguage {en}\bibinfo {title} {Damping of relaxation oscillations, photon-photon resonance, and tolerance to external optical feedback of {III}-{V}/{SiN} hybrid lasers with a dispersive narrow band mirror}},\ }\href {https://doi.org/10.1364/OE.452155} {\bibfield  {journal} {\bibinfo  {journal} {Optics Express}\ }\textbf {\bibinfo {volume} {30}},\ \bibinfo {pages} {11090} (\bibinfo {year} {2022})}\BibitemShut {NoStop}%
\bibitem [{\citenamefont {Ramunno}\ and\ \citenamefont {Sipe}(2002)}]{ramunno_stability_2002}%
  \BibitemOpen
  \bibfield  {author} {\bibinfo {author} {\bibfnamefont {L.}~\bibnamefont {Ramunno}}\ and\ \bibinfo {author} {\bibfnamefont {J.~E.}\ \bibnamefont {Sipe}},\ }\bibfield  {title} {{\selectlanguage {en}\bibinfo {title} {Stability of a semiconductor laser with a dispersive extended cavity}},\ }\href {https://doi.org/10.1103/PhysRevA.66.033817} {\bibfield  {journal} {\bibinfo  {journal} {Physical Review A}\ }\textbf {\bibinfo {volume} {66}},\ \bibinfo {pages} {033817} (\bibinfo {year} {2002})}\BibitemShut {NoStop}%
\bibitem [{\citenamefont {Mak}\ \emph {et~al.}(2019)\citenamefont {Mak}, \citenamefont {van Rees}, \citenamefont {Fan}, \citenamefont {Klein}, \citenamefont {Geskus}, \citenamefont {van~der Slot},\ and\ \citenamefont {Boller}}]{mak_linewidth_2019}%
  \BibitemOpen
  \bibfield  {author} {\bibinfo {author} {\bibfnamefont {J.}~\bibnamefont {Mak}}, \bibinfo {author} {\bibfnamefont {A.}~\bibnamefont {van Rees}}, \bibinfo {author} {\bibfnamefont {Y.}~\bibnamefont {Fan}}, \bibinfo {author} {\bibfnamefont {E.~J.}\ \bibnamefont {Klein}}, \bibinfo {author} {\bibfnamefont {D.}~\bibnamefont {Geskus}}, \bibinfo {author} {\bibfnamefont {P.~J.~M.}\ \bibnamefont {van~der Slot}},\ and\ \bibinfo {author} {\bibfnamefont {K.-J.}\ \bibnamefont {Boller}},\ }\bibfield  {title} {{\selectlanguage {en}\bibinfo {title} {Linewidth narrowing via low-loss dielectric waveguide feedback circuits in hybrid integrated frequency comb lasers}},\ }\href {https://doi.org/10.1364/OE.27.013307} {\bibfield  {journal} {\bibinfo  {journal} {Optics Express}\ }\textbf {\bibinfo {volume} {27}},\ \bibinfo {pages} {13307} (\bibinfo {year} {2019})}\BibitemShut {NoStop}%
\bibitem [{\citenamefont {Bandelow}\ \emph {et~al.}(2000)\citenamefont {Bandelow}, \citenamefont {Radziunas}, \citenamefont {Tronciu}, \citenamefont {Wunsche},\ and\ \citenamefont {Henneberger}}]{bandelow_tailoring_2000}%
  \BibitemOpen
  \bibfield  {author} {\bibinfo {author} {\bibfnamefont {U.}~\bibnamefont {Bandelow}}, \bibinfo {author} {\bibfnamefont {M.}~\bibnamefont {Radziunas}}, \bibinfo {author} {\bibfnamefont {V.~Z.}\ \bibnamefont {Tronciu}}, \bibinfo {author} {\bibfnamefont {H.-J.}\ \bibnamefont {Wunsche}},\ and\ \bibinfo {author} {\bibfnamefont {F.}~\bibnamefont {Henneberger}},\ }\bibfield  {title} {\bibinfo {title} {Tailoring the dynamics of diode lasers by passive dispersive reflectors}\ }(\bibinfo {address} {San Jose, CA},\ \bibinfo {year} {2000})\ p.\ \bibinfo {pages} {536}\BibitemShut {NoStop}%
\bibitem [{\citenamefont {Yu}\ \emph {et~al.}(2022)\citenamefont {Yu}, \citenamefont {Zali},\ and\ \citenamefont {Mørk}}]{yu_theory_2022}%
  \BibitemOpen
  \bibfield  {author} {\bibinfo {author} {\bibfnamefont {Y.}~\bibnamefont {Yu}}, \bibinfo {author} {\bibfnamefont {A.~R.}\ \bibnamefont {Zali}},\ and\ \bibinfo {author} {\bibfnamefont {J.}~\bibnamefont {Mørk}},\ }\bibfield  {title} {{\selectlanguage {en}\bibinfo {title} {Theory of linewidth narrowing in {Fano} lasers}},\ }\href {https://doi.org/10.1103/PhysRevResearch.4.043194} {\bibfield  {journal} {\bibinfo  {journal} {Physical Review Research}\ }\textbf {\bibinfo {volume} {4}},\ \bibinfo {pages} {043194} (\bibinfo {year} {2022})}\BibitemShut {NoStop}%
\bibitem [{\citenamefont {Mork}\ \emph {et~al.}(2014)\citenamefont {Mork}, \citenamefont {Chen},\ and\ \citenamefont {Heuck}}]{mork_photonic_2014}%
  \BibitemOpen
  \bibfield  {author} {\bibinfo {author} {\bibfnamefont {J.}~\bibnamefont {Mork}}, \bibinfo {author} {\bibfnamefont {Y.}~\bibnamefont {Chen}},\ and\ \bibinfo {author} {\bibfnamefont {M.}~\bibnamefont {Heuck}},\ }\bibfield  {title} {{\selectlanguage {en}\bibinfo {title} {Photonic {Crystal} {Fano} {Laser}: {Terahertz} {Modulation} and {Ultrashort} {Pulse} {Generation}}},\ }\href {https://doi.org/10.1103/PhysRevLett.113.163901} {\bibfield  {journal} {\bibinfo  {journal} {Physical Review Letters}\ }\textbf {\bibinfo {volume} {113}},\ \bibinfo {pages} {163901} (\bibinfo {year} {2014})}\BibitemShut {NoStop}%
\bibitem [{\citenamefont {Mork}\ \emph {et~al.}(2019)\citenamefont {Mork}, \citenamefont {Yu}, \citenamefont {Rasmussen}, \citenamefont {Semenova},\ and\ \citenamefont {Yvind}}]{mork_semiconductor_2019}%
  \BibitemOpen
  \bibfield  {author} {\bibinfo {author} {\bibfnamefont {J.}~\bibnamefont {Mork}}, \bibinfo {author} {\bibfnamefont {Y.}~\bibnamefont {Yu}}, \bibinfo {author} {\bibfnamefont {T.~S.}\ \bibnamefont {Rasmussen}}, \bibinfo {author} {\bibfnamefont {E.}~\bibnamefont {Semenova}},\ and\ \bibinfo {author} {\bibfnamefont {K.}~\bibnamefont {Yvind}},\ }\bibfield  {title} {\bibinfo {title} {Semiconductor {Fano} {Lasers}},\ }\href {https://doi.org/10.1109/JSTQE.2019.2922067} {\bibfield  {journal} {\bibinfo  {journal} {IEEE Journal of Selected Topics in Quantum Electronics}\ }\textbf {\bibinfo {volume} {25}},\ \bibinfo {pages} {1} (\bibinfo {year} {2019})}\BibitemShut {NoStop}%
\bibitem [{\citenamefont {Fano}(1961)}]{fano_effects_1961}%
  \BibitemOpen
  \bibfield  {author} {\bibinfo {author} {\bibfnamefont {U.}~\bibnamefont {Fano}},\ }\bibfield  {title} {{\selectlanguage {en}\bibinfo {title} {Effects of {Configuration} {Interaction} on {Intensities} and {Phase} {Shifts}}},\ }\href {https://doi.org/10.1103/PhysRev.124.1866} {\bibfield  {journal} {\bibinfo  {journal} {Physical Review}\ }\textbf {\bibinfo {volume} {124}},\ \bibinfo {pages} {1866} (\bibinfo {year} {1961})}\BibitemShut {NoStop}%
\bibitem [{\citenamefont {Matsuo}\ \emph {et~al.}(2010)\citenamefont {Matsuo}, \citenamefont {Shinya}, \citenamefont {Kakitsuka}, \citenamefont {Nozaki}, \citenamefont {Segawa}, \citenamefont {Sato}, \citenamefont {Kawaguchi},\ and\ \citenamefont {Notomi}}]{matsuo_high-speed_2010}%
  \BibitemOpen
  \bibfield  {author} {\bibinfo {author} {\bibfnamefont {S.}~\bibnamefont {Matsuo}}, \bibinfo {author} {\bibfnamefont {A.}~\bibnamefont {Shinya}}, \bibinfo {author} {\bibfnamefont {T.}~\bibnamefont {Kakitsuka}}, \bibinfo {author} {\bibfnamefont {K.}~\bibnamefont {Nozaki}}, \bibinfo {author} {\bibfnamefont {T.}~\bibnamefont {Segawa}}, \bibinfo {author} {\bibfnamefont {T.}~\bibnamefont {Sato}}, \bibinfo {author} {\bibfnamefont {Y.}~\bibnamefont {Kawaguchi}},\ and\ \bibinfo {author} {\bibfnamefont {M.}~\bibnamefont {Notomi}},\ }\bibfield  {title} {{\selectlanguage {en}\bibinfo {title} {High-speed ultracompact buried heterostructure photonic-crystal laser with 13 {fJ} of energy consumed per bit transmitted}},\ }\href {https://doi.org/10.1038/nphoton.2010.177} {\bibfield  {journal} {\bibinfo  {journal} {Nature Photonics}\ }\textbf {\bibinfo {volume} {4}},\ \bibinfo {pages} {648} (\bibinfo {year} {2010})}\BibitemShut {NoStop}%
\bibitem [{\citenamefont {Yu}\ \emph {et~al.}(2021)\citenamefont {Yu}, \citenamefont {Sakanas}, \citenamefont {Zali}, \citenamefont {Semenova}, \citenamefont {Yvind},\ and\ \citenamefont {Mørk}}]{yu_ultra-coherent_2021}%
  \BibitemOpen
  \bibfield  {author} {\bibinfo {author} {\bibfnamefont {Y.}~\bibnamefont {Yu}}, \bibinfo {author} {\bibfnamefont {A.}~\bibnamefont {Sakanas}}, \bibinfo {author} {\bibfnamefont {A.~R.}\ \bibnamefont {Zali}}, \bibinfo {author} {\bibfnamefont {E.}~\bibnamefont {Semenova}}, \bibinfo {author} {\bibfnamefont {K.}~\bibnamefont {Yvind}},\ and\ \bibinfo {author} {\bibfnamefont {J.}~\bibnamefont {Mørk}},\ }\bibfield  {title} {{\selectlanguage {en}\bibinfo {title} {Ultra-coherent {Fano} laser based on a bound state in the continuum}},\ }\href {https://doi.org/10.1038/s41566-021-00860-5} {\bibfield  {journal} {\bibinfo  {journal} {Nature Photonics}\ }\textbf {\bibinfo {volume} {15}},\ \bibinfo {pages} {758} (\bibinfo {year} {2021})}\BibitemShut {NoStop}%
\bibitem [{\citenamefont {Fan}\ \emph {et~al.}(2003)\citenamefont {Fan}, \citenamefont {Suh},\ and\ \citenamefont {Joannopoulos}}]{fan_temporal_2003}%
  \BibitemOpen
  \bibfield  {author} {\bibinfo {author} {\bibfnamefont {S.}~\bibnamefont {Fan}}, \bibinfo {author} {\bibfnamefont {W.}~\bibnamefont {Suh}},\ and\ \bibinfo {author} {\bibfnamefont {J.~D.}\ \bibnamefont {Joannopoulos}},\ }\bibfield  {title} {{\selectlanguage {en}\bibinfo {title} {Temporal coupled-mode theory for the {Fano} resonance in optical resonators}},\ }\href {https://doi.org/10.1364/JOSAA.20.000569} {\bibfield  {journal} {\bibinfo  {journal} {Journal of the Optical Society of America A}\ }\textbf {\bibinfo {volume} {20}},\ \bibinfo {pages} {569} (\bibinfo {year} {2003})}\BibitemShut {NoStop}%
\bibitem [{\citenamefont {Rasmussen}\ \emph {et~al.}(2019)\citenamefont {Rasmussen}, \citenamefont {Yu},\ and\ \citenamefont {Mork}}]{rasmussen_suppression_2019}%
  \BibitemOpen
  \bibfield  {author} {\bibinfo {author} {\bibfnamefont {T.~S.}\ \bibnamefont {Rasmussen}}, \bibinfo {author} {\bibfnamefont {Y.}~\bibnamefont {Yu}},\ and\ \bibinfo {author} {\bibfnamefont {J.}~\bibnamefont {Mork}},\ }\bibfield  {title} {{\selectlanguage {en}\bibinfo {title} {Suppression of {Coherence} {Collapse} in {Semiconductor} {Fano} {Lasers}}},\ }\href {https://doi.org/10.1103/PhysRevLett.123.233904} {\bibfield  {journal} {\bibinfo  {journal} {Physical Review Letters}\ }\textbf {\bibinfo {volume} {123}},\ \bibinfo {pages} {233904} (\bibinfo {year} {2019})}\BibitemShut {NoStop}%
\bibitem [{\citenamefont {Dong}\ \emph {et~al.}(2023)\citenamefont {Dong}, \citenamefont {Liang}, \citenamefont {Sakanas}, \citenamefont {Semenova}, \citenamefont {Yvind}, \citenamefont {Mørk},\ and\ \citenamefont {Yu}}]{dong_cavity_2023}%
  \BibitemOpen
  \bibfield  {author} {\bibinfo {author} {\bibfnamefont {G.}~\bibnamefont {Dong}}, \bibinfo {author} {\bibfnamefont {S.~L.}\ \bibnamefont {Liang}}, \bibinfo {author} {\bibfnamefont {A.}~\bibnamefont {Sakanas}}, \bibinfo {author} {\bibfnamefont {E.}~\bibnamefont {Semenova}}, \bibinfo {author} {\bibfnamefont {K.}~\bibnamefont {Yvind}}, \bibinfo {author} {\bibfnamefont {J.}~\bibnamefont {Mørk}},\ and\ \bibinfo {author} {\bibfnamefont {Y.}~\bibnamefont {Yu}},\ }\bibfield  {title} {{\selectlanguage {en}\bibinfo {title} {Cavity dumping using a microscopic {Fano} laser}},\ }\href {https://doi.org/10.1364/OPTICA.476758} {\bibfield  {journal} {\bibinfo  {journal} {Optica}\ }\textbf {\bibinfo {volume} {10}},\ \bibinfo {pages} {248} (\bibinfo {year} {2023})}\BibitemShut {NoStop}%
\bibitem [{\citenamefont {Bekele}\ \emph {et~al.}(2019)\citenamefont {Bekele}, \citenamefont {Yu}, \citenamefont {Yvind},\ and\ \citenamefont {Mork}}]{bekele_plane_2019}%
  \BibitemOpen
  \bibfield  {author} {\bibinfo {author} {\bibfnamefont {D.}~\bibnamefont {Bekele}}, \bibinfo {author} {\bibfnamefont {Y.}~\bibnamefont {Yu}}, \bibinfo {author} {\bibfnamefont {K.}~\bibnamefont {Yvind}},\ and\ \bibinfo {author} {\bibfnamefont {J.}~\bibnamefont {Mork}},\ }\bibfield  {title} {{\selectlanguage {en}\bibinfo {title} {In‐{Plane} {Photonic} {Crystal} {Devices} using {Fano} {Resonances}}},\ }\href {https://doi.org/10.1002/lpor.201900054} {\bibfield  {journal} {\bibinfo  {journal} {Laser \& Photonics Reviews}\ }\textbf {\bibinfo {volume} {13}},\ \bibinfo {pages} {1900054} (\bibinfo {year} {2019})}\BibitemShut {NoStop}%
\bibitem [{\citenamefont {{Wonjoo Suh}}\ \emph {et~al.}(2004)\citenamefont {{Wonjoo Suh}}, \citenamefont {{Zheng Wang}},\ and\ \citenamefont {{Shanhui Fan}}}]{wonjoo_suh_temporal_2004}%
  \BibitemOpen
  \bibfield  {author} {\bibinfo {author} {\bibnamefont {{Wonjoo Suh}}}, \bibinfo {author} {\bibnamefont {{Zheng Wang}}},\ and\ \bibinfo {author} {\bibnamefont {{Shanhui Fan}}},\ }\bibfield  {title} {\bibinfo {title} {Temporal coupled-mode theory and the presence of non-orthogonal modes in lossless multimode cavities},\ }\href {https://doi.org/10.1109/JQE.2004.834773} {\bibfield  {journal} {\bibinfo  {journal} {IEEE Journal of Quantum Electronics}\ }\textbf {\bibinfo {volume} {40}},\ \bibinfo {pages} {1511} (\bibinfo {year} {2004})}\BibitemShut {NoStop}%
\bibitem [{\citenamefont {Yan}\ \emph {et~al.}(2023)\citenamefont {Yan}, \citenamefont {Jiang}, \citenamefont {Li},\ and\ \citenamefont {Deng}}]{yan_controlling_2023}%
  \BibitemOpen
  \bibfield  {author} {\bibinfo {author} {\bibfnamefont {Y.}~\bibnamefont {Yan}}, \bibinfo {author} {\bibfnamefont {Y.-F.}\ \bibnamefont {Jiang}}, \bibinfo {author} {\bibfnamefont {B.-X.}\ \bibnamefont {Li}},\ and\ \bibinfo {author} {\bibfnamefont {C.-S.}\ \bibnamefont {Deng}},\ }\bibfield  {title} {\bibinfo {title} {Controlling {Dual} {Fano} {Resonance} {Lineshapes} {Based} on an {Indirectly} {Coupled} {Double}-{Nanobeam}-{Cavity} {Photonic} {Molecule}},\ }\href {https://doi.org/10.1109/JLT.2023.3318291} {\bibfield  {journal} {\bibinfo  {journal} {Journal of Lightwave Technology}\ ,\ \bibinfo {pages} {1}} (\bibinfo {year} {2023})}\BibitemShut {NoStop}%
\bibitem [{\citenamefont {Heuck}\ \emph {et~al.}(2014)\citenamefont {Heuck}, \citenamefont {Kristensen},\ and\ \citenamefont {Mørk}}]{heuck_dual-resonances_2014}%
  \BibitemOpen
  \bibfield  {author} {\bibinfo {author} {\bibfnamefont {M.}~\bibnamefont {Heuck}}, \bibinfo {author} {\bibfnamefont {P.~T.}\ \bibnamefont {Kristensen}},\ and\ \bibinfo {author} {\bibfnamefont {J.}~\bibnamefont {Mørk}},\ }\bibfield  {title} {{\selectlanguage {en}\bibinfo {title} {Dual-resonances approach to broadband cavity-assisted optical signal processing beyond the carrier relaxation rate}},\ }\href {https://doi.org/10.1364/OL.39.003189} {\bibfield  {journal} {\bibinfo  {journal} {Optics Letters}\ }\textbf {\bibinfo {volume} {39}},\ \bibinfo {pages} {3189} (\bibinfo {year} {2014})}\BibitemShut {NoStop}%
\bibitem [{\citenamefont {Hsu}\ \emph {et~al.}(2016)\citenamefont {Hsu}, \citenamefont {Zhen}, \citenamefont {Stone}, \citenamefont {Joannopoulos},\ and\ \citenamefont {Soljačić}}]{hsu_bound_2016}%
  \BibitemOpen
  \bibfield  {author} {\bibinfo {author} {\bibfnamefont {C.~W.}\ \bibnamefont {Hsu}}, \bibinfo {author} {\bibfnamefont {B.}~\bibnamefont {Zhen}}, \bibinfo {author} {\bibfnamefont {A.~D.}\ \bibnamefont {Stone}}, \bibinfo {author} {\bibfnamefont {J.~D.}\ \bibnamefont {Joannopoulos}},\ and\ \bibinfo {author} {\bibfnamefont {M.}~\bibnamefont {Soljačić}},\ }\bibfield  {title} {{\selectlanguage {en}\bibinfo {title} {Bound states in the continuum}},\ }\href {https://doi.org/10.1038/natrevmats.2016.48} {\bibfield  {journal} {\bibinfo  {journal} {Nature Reviews Materials}\ }\textbf {\bibinfo {volume} {1}},\ \bibinfo {pages} {16048} (\bibinfo {year} {2016})}\BibitemShut {NoStop}%
\bibitem [{\citenamefont {Kristensen}\ \emph {et~al.}(2017)\citenamefont {Kristensen}, \citenamefont {de~Lasson}, \citenamefont {Heuck}, \citenamefont {Gregersen},\ and\ \citenamefont {Mork}}]{kristensen_theory_2017}%
  \BibitemOpen
  \bibfield  {author} {\bibinfo {author} {\bibfnamefont {P.~T.}\ \bibnamefont {Kristensen}}, \bibinfo {author} {\bibfnamefont {J.~R.}\ \bibnamefont {de~Lasson}}, \bibinfo {author} {\bibfnamefont {M.}~\bibnamefont {Heuck}}, \bibinfo {author} {\bibfnamefont {N.}~\bibnamefont {Gregersen}},\ and\ \bibinfo {author} {\bibfnamefont {J.}~\bibnamefont {Mork}},\ }\bibfield  {title} {\bibinfo {title} {On the {Theory} of {Coupled} {Modes} in {Optical} {Cavity}-{Waveguide} {Structures}},\ }\href {https://doi.org/10.1109/JLT.2017.2714263} {\bibfield  {journal} {\bibinfo  {journal} {Journal of Lightwave Technology}\ }\textbf {\bibinfo {volume} {35}},\ \bibinfo {pages} {4247} (\bibinfo {year} {2017})}\BibitemShut {NoStop}%
\bibitem [{\citenamefont {Chalcraft}\ \emph {et~al.}(2011)\citenamefont {Chalcraft}, \citenamefont {Lam}, \citenamefont {Jones}, \citenamefont {Szymanski}, \citenamefont {Oulton}, \citenamefont {Thijssen}, \citenamefont {Skolnick}, \citenamefont {Whittaker}, \citenamefont {Krauss},\ and\ \citenamefont {Fox}}]{chalcraft_mode_2011}%
  \BibitemOpen
  \bibfield  {author} {\bibinfo {author} {\bibfnamefont {A.~R.~A.}\ \bibnamefont {Chalcraft}}, \bibinfo {author} {\bibfnamefont {S.}~\bibnamefont {Lam}}, \bibinfo {author} {\bibfnamefont {B.~D.}\ \bibnamefont {Jones}}, \bibinfo {author} {\bibfnamefont {D.}~\bibnamefont {Szymanski}}, \bibinfo {author} {\bibfnamefont {R.}~\bibnamefont {Oulton}}, \bibinfo {author} {\bibfnamefont {A.~C.~T.}\ \bibnamefont {Thijssen}}, \bibinfo {author} {\bibfnamefont {M.~S.}\ \bibnamefont {Skolnick}}, \bibinfo {author} {\bibfnamefont {D.~M.}\ \bibnamefont {Whittaker}}, \bibinfo {author} {\bibfnamefont {T.~F.}\ \bibnamefont {Krauss}},\ and\ \bibinfo {author} {\bibfnamefont {A.~M.}\ \bibnamefont {Fox}},\ }\bibfield  {title} {{\selectlanguage {en}\bibinfo {title} {Mode structure of coupled {L3} photonic crystal cavities}},\ }\href {https://doi.org/10.1364/OE.19.005670} {\bibfield  {journal} {\bibinfo  {journal} {Optics Express}\ }\textbf {\bibinfo {volume} {19}},\ \bibinfo {pages} {5670} (\bibinfo {year} {2011})}\BibitemShut
  {NoStop}%
\bibitem [{\citenamefont {Haddadi}\ \emph {et~al.}(2014)\citenamefont {Haddadi}, \citenamefont {Hamel}, \citenamefont {Beaudoin}, \citenamefont {Sagnes}, \citenamefont {Sauvan}, \citenamefont {Lalanne}, \citenamefont {Levenson},\ and\ \citenamefont {Yacomotti}}]{haddadi_photonic_2014}%
  \BibitemOpen
  \bibfield  {author} {\bibinfo {author} {\bibfnamefont {S.}~\bibnamefont {Haddadi}}, \bibinfo {author} {\bibfnamefont {P.}~\bibnamefont {Hamel}}, \bibinfo {author} {\bibfnamefont {G.}~\bibnamefont {Beaudoin}}, \bibinfo {author} {\bibfnamefont {I.}~\bibnamefont {Sagnes}}, \bibinfo {author} {\bibfnamefont {C.}~\bibnamefont {Sauvan}}, \bibinfo {author} {\bibfnamefont {P.}~\bibnamefont {Lalanne}}, \bibinfo {author} {\bibfnamefont {J.~A.}\ \bibnamefont {Levenson}},\ and\ \bibinfo {author} {\bibfnamefont {A.~M.}\ \bibnamefont {Yacomotti}},\ }\bibfield  {title} {{\selectlanguage {en}\bibinfo {title} {Photonic molecules: tailoring the coupling strength and sign}},\ }\href {https://doi.org/10.1364/OE.22.012359} {\bibfield  {journal} {\bibinfo  {journal} {Optics Express}\ }\textbf {\bibinfo {volume} {22}},\ \bibinfo {pages} {12359} (\bibinfo {year} {2014})}\BibitemShut {NoStop}%
\bibitem [{\citenamefont {Yu}\ \emph {et~al.}(2013)\citenamefont {Yu}, \citenamefont {Palushani}, \citenamefont {Heuck}, \citenamefont {Kuznetsova}, \citenamefont {Kristensen}, \citenamefont {Ek}, \citenamefont {Vukovic}, \citenamefont {Peucheret}, \citenamefont {Oxenløwe}, \citenamefont {Combrié}, \citenamefont {de~Rossi}, \citenamefont {Yvind},\ and\ \citenamefont {Mørk}}]{yu_switching_2013}%
  \BibitemOpen
  \bibfield  {author} {\bibinfo {author} {\bibfnamefont {Y.}~\bibnamefont {Yu}}, \bibinfo {author} {\bibfnamefont {E.}~\bibnamefont {Palushani}}, \bibinfo {author} {\bibfnamefont {M.}~\bibnamefont {Heuck}}, \bibinfo {author} {\bibfnamefont {N.}~\bibnamefont {Kuznetsova}}, \bibinfo {author} {\bibfnamefont {P.~T.}\ \bibnamefont {Kristensen}}, \bibinfo {author} {\bibfnamefont {S.}~\bibnamefont {Ek}}, \bibinfo {author} {\bibfnamefont {D.}~\bibnamefont {Vukovic}}, \bibinfo {author} {\bibfnamefont {C.}~\bibnamefont {Peucheret}}, \bibinfo {author} {\bibfnamefont {L.~K.}\ \bibnamefont {Oxenløwe}}, \bibinfo {author} {\bibfnamefont {S.}~\bibnamefont {Combrié}}, \bibinfo {author} {\bibfnamefont {A.}~\bibnamefont {de~Rossi}}, \bibinfo {author} {\bibfnamefont {K.}~\bibnamefont {Yvind}},\ and\ \bibinfo {author} {\bibfnamefont {J.}~\bibnamefont {Mørk}},\ }\bibfield  {title} {{\selectlanguage {en}\bibinfo {title} {Switching characteristics of an {InP} photonic crystal nanocavity: {Experiment} and theory}},\ }\href
  {https://doi.org/10.1364/OE.21.031047} {\bibfield  {journal} {\bibinfo  {journal} {Optics Express}\ }\textbf {\bibinfo {volume} {21}},\ \bibinfo {pages} {31047} (\bibinfo {year} {2013})}\BibitemShut {NoStop}%
\bibitem [{\citenamefont {Detoma}\ \emph {et~al.}(2005)\citenamefont {Detoma}, \citenamefont {Tromborg},\ and\ \citenamefont {Montrosset}}]{detoma_complex_2005}%
  \BibitemOpen
  \bibfield  {author} {\bibinfo {author} {\bibfnamefont {E.}~\bibnamefont {Detoma}}, \bibinfo {author} {\bibfnamefont {B.}~\bibnamefont {Tromborg}},\ and\ \bibinfo {author} {\bibfnamefont {I.}~\bibnamefont {Montrosset}},\ }\bibfield  {title} {{\selectlanguage {en}\bibinfo {title} {The complex way to laser diode spectra: example of an external cavity laser strong optical feedback}},\ }\href {https://doi.org/10.1109/JQE.2004.839705} {\bibfield  {journal} {\bibinfo  {journal} {IEEE Journal of Quantum Electronics}\ }\textbf {\bibinfo {volume} {41}},\ \bibinfo {pages} {171} (\bibinfo {year} {2005})}\BibitemShut {NoStop}%
\bibitem [{\citenamefont {Piprek}(2005)}]{piprek_optoelectronic_2005}%
  \BibitemOpen
  \bibfield  {author} {\bibinfo {author} {\bibfnamefont {J.}~\bibnamefont {Piprek}},\ }\href@noop {} {{\selectlanguage {eng}\emph {\bibinfo {title} {Optoelectronic devices: advanced simulation and analysis}}}}\ (\bibinfo  {publisher} {Springer},\ \bibinfo {address} {New York},\ \bibinfo {year} {2005})\BibitemShut {NoStop}%
\bibitem [{\citenamefont {Tromborg}\ \emph {et~al.}(1997)\citenamefont {Tromborg}, \citenamefont {Mørk},\ and\ \citenamefont {Velichansky}}]{tromborg_mode_1997}%
  \BibitemOpen
  \bibfield  {author} {\bibinfo {author} {\bibfnamefont {B.}~\bibnamefont {Tromborg}}, \bibinfo {author} {\bibfnamefont {J.}~\bibnamefont {Mørk}},\ and\ \bibinfo {author} {\bibfnamefont {V.}~\bibnamefont {Velichansky}},\ }\bibfield  {title} {{\selectlanguage {en}\bibinfo {title} {On mode coupling and low-frequency fluctuations in external-cavity laser diodes}},\ }\href {https://doi.org/10.1088/1355-5111/9/5/014} {\bibfield  {journal} {\bibinfo  {journal} {Quantum and Semiclassical Optics: Journal of the European Optical Society Part B}\ }\textbf {\bibinfo {volume} {9}},\ \bibinfo {pages} {831} (\bibinfo {year} {1997})}\BibitemShut {NoStop}%
\bibitem [{\citenamefont {Bogatov}\ \emph {et~al.}(1975)\citenamefont {Bogatov}, \citenamefont {Eliseev},\ and\ \citenamefont {Sverdlov}}]{bogatov_anomalous_1975}%
  \BibitemOpen
  \bibfield  {author} {\bibinfo {author} {\bibfnamefont {A.}~\bibnamefont {Bogatov}}, \bibinfo {author} {\bibfnamefont {P.}~\bibnamefont {Eliseev}},\ and\ \bibinfo {author} {\bibfnamefont {B.}~\bibnamefont {Sverdlov}},\ }\bibfield  {title} {\bibinfo {title} {Anomalous interaction of spectral modes in a semiconductor laser},\ }\href {https://doi.org/10.1109/JQE.1975.1068658} {\bibfield  {journal} {\bibinfo  {journal} {IEEE Journal of Quantum Electronics}\ }\textbf {\bibinfo {volume} {11}},\ \bibinfo {pages} {510} (\bibinfo {year} {1975})}\BibitemShut {NoStop}%
\bibitem [{\citenamefont {Tronciu}\ \emph {et~al.}(2021)\citenamefont {Tronciu}, \citenamefont {Werner}, \citenamefont {Wenzel},\ and\ \citenamefont {Wunsche}}]{tronciu_feedback_2021}%
  \BibitemOpen
  \bibfield  {author} {\bibinfo {author} {\bibfnamefont {V.}~\bibnamefont {Tronciu}}, \bibinfo {author} {\bibfnamefont {N.}~\bibnamefont {Werner}}, \bibinfo {author} {\bibfnamefont {H.}~\bibnamefont {Wenzel}},\ and\ \bibinfo {author} {\bibfnamefont {H.-J.}\ \bibnamefont {Wunsche}},\ }\bibfield  {title} {{\selectlanguage {en}\bibinfo {title} {Feedback {Sensitivity} of {Detuned} {DBR} {Semiconductor} {Lasers}}},\ }\href {https://doi.org/10.1109/JQE.2021.3101216} {\bibfield  {journal} {\bibinfo  {journal} {IEEE Journal of Quantum Electronics}\ }\textbf {\bibinfo {volume} {57}},\ \bibinfo {pages} {1} (\bibinfo {year} {2021})}\BibitemShut {NoStop}%
\bibitem [{\citenamefont {Vahala}\ and\ \citenamefont {Yariv}(1984)}]{vahala_detuned_1984}%
  \BibitemOpen
  \bibfield  {author} {\bibinfo {author} {\bibfnamefont {K.}~\bibnamefont {Vahala}}\ and\ \bibinfo {author} {\bibfnamefont {A.}~\bibnamefont {Yariv}},\ }\bibfield  {title} {{\selectlanguage {en}\bibinfo {title} {Detuned loading in coupled cavity semiconductor lasers—effect on quantum noise and dynamics}},\ }\href {https://doi.org/10.1063/1.95316} {\bibfield  {journal} {\bibinfo  {journal} {Applied Physics Letters}\ }\textbf {\bibinfo {volume} {45}},\ \bibinfo {pages} {501} (\bibinfo {year} {1984})}\BibitemShut {NoStop}%
\bibitem [{\citenamefont {Tromborg}\ \emph {et~al.}(1987)\citenamefont {Tromborg}, \citenamefont {Olesen}, \citenamefont {{Xing Pan}},\ and\ \citenamefont {Saito}}]{tromborg_transmission_1987}%
  \BibitemOpen
  \bibfield  {author} {\bibinfo {author} {\bibfnamefont {B.}~\bibnamefont {Tromborg}}, \bibinfo {author} {\bibfnamefont {H.}~\bibnamefont {Olesen}}, \bibinfo {author} {\bibnamefont {{Xing Pan}}},\ and\ \bibinfo {author} {\bibfnamefont {S.}~\bibnamefont {Saito}},\ }\bibfield  {title} {{\selectlanguage {en}\bibinfo {title} {Transmission line description of optical feedback and injection locking for {Fabry}-{Perot} and {DFB} lasers}},\ }\href {https://doi.org/10.1109/JQE.1987.1073251} {\bibfield  {journal} {\bibinfo  {journal} {IEEE Journal of Quantum Electronics}\ }\textbf {\bibinfo {volume} {23}},\ \bibinfo {pages} {1875} (\bibinfo {year} {1987})}\BibitemShut {NoStop}%
\bibitem [{\citenamefont {Coldren}\ \emph {et~al.}(2012)\citenamefont {Coldren}, \citenamefont {Corzine},\ and\ \citenamefont {Mašanović}}]{coldren_diode_2012}%
  \BibitemOpen
  \bibfield  {author} {\bibinfo {author} {\bibfnamefont {L.~A.}\ \bibnamefont {Coldren}}, \bibinfo {author} {\bibfnamefont {S.~W.}\ \bibnamefont {Corzine}},\ and\ \bibinfo {author} {\bibfnamefont {M.~L.}\ \bibnamefont {Mašanović}},\ }\href {https://doi.org/10.1002/9781118148167} {\emph {\bibinfo {title} {Diode {Lasers} and {Photonic} {Integrated} {Circuits}: {Coldren}/{Diode} {Lasers} {2E}}}}\ (\bibinfo  {publisher} {John Wiley \& Sons, Inc.},\ \bibinfo {address} {Hoboken, NJ, USA},\ \bibinfo {year} {2012})\BibitemShut {NoStop}%
\bibitem [{\citenamefont {Rasmussen}\ \emph {et~al.}(2018)\citenamefont {Rasmussen}, \citenamefont {Yu},\ and\ \citenamefont {Mork}}]{rasmussen_modes_2018}%
  \BibitemOpen
  \bibfield  {author} {\bibinfo {author} {\bibfnamefont {T.~S.}\ \bibnamefont {Rasmussen}}, \bibinfo {author} {\bibfnamefont {Y.}~\bibnamefont {Yu}},\ and\ \bibinfo {author} {\bibfnamefont {J.}~\bibnamefont {Mork}},\ }\bibfield  {title} {{\selectlanguage {en}\bibinfo {title} {Modes, stability, and small-signal response of photonic crystal {Fano} lasers}},\ }\href {https://doi.org/10.1364/OE.26.016365} {\bibfield  {journal} {\bibinfo  {journal} {Optics Express}\ }\textbf {\bibinfo {volume} {26}},\ \bibinfo {pages} {16365} (\bibinfo {year} {2018})}\BibitemShut {NoStop}%
\bibitem [{\citenamefont {Agrawal}\ and\ \citenamefont {Dutta}(1993)}]{agrawal_semiconductor_1993}%
  \BibitemOpen
  \bibfield  {author} {\bibinfo {author} {\bibfnamefont {G.~P.}\ \bibnamefont {Agrawal}}\ and\ \bibinfo {author} {\bibfnamefont {N.~K.}\ \bibnamefont {Dutta}},\ }\href {https://doi.org/10.1007/978-1-4613-0481-4} {{\selectlanguage {en}\emph {\bibinfo {title} {Semiconductor {Lasers}}}}}\ (\bibinfo  {publisher} {Springer US},\ \bibinfo {address} {Boston, MA},\ \bibinfo {year} {1993})\BibitemShut {NoStop}%
\bibitem [{\citenamefont {Marconi}\ \emph {et~al.}(2016)\citenamefont {Marconi}, \citenamefont {Javaloyes}, \citenamefont {Raineri}, \citenamefont {Levenson},\ and\ \citenamefont {Yacomotti}}]{marconi_asymmetric_2016}%
  \BibitemOpen
  \bibfield  {author} {\bibinfo {author} {\bibfnamefont {M.}~\bibnamefont {Marconi}}, \bibinfo {author} {\bibfnamefont {J.}~\bibnamefont {Javaloyes}}, \bibinfo {author} {\bibfnamefont {F.}~\bibnamefont {Raineri}}, \bibinfo {author} {\bibfnamefont {J.~A.}\ \bibnamefont {Levenson}},\ and\ \bibinfo {author} {\bibfnamefont {A.~M.}\ \bibnamefont {Yacomotti}},\ }\bibfield  {title} {{\selectlanguage {en}\bibinfo {title} {Asymmetric mode scattering in strongly coupled photonic crystal nanolasers}},\ }\href {https://doi.org/10.1364/OL.41.005628} {\bibfield  {journal} {\bibinfo  {journal} {Optics Letters}\ }\textbf {\bibinfo {volume} {41}},\ \bibinfo {pages} {5628} (\bibinfo {year} {2016})}\BibitemShut {NoStop}%
\bibitem [{\citenamefont {Wünsche}\ \emph {et~al.}(2001)\citenamefont {Wünsche}, \citenamefont {Brox}, \citenamefont {Radziunas},\ and\ \citenamefont {Henneberger}}]{wunsche_excitability_2001}%
  \BibitemOpen
  \bibfield  {author} {\bibinfo {author} {\bibfnamefont {H.~J.}\ \bibnamefont {Wünsche}}, \bibinfo {author} {\bibfnamefont {O.}~\bibnamefont {Brox}}, \bibinfo {author} {\bibfnamefont {M.}~\bibnamefont {Radziunas}},\ and\ \bibinfo {author} {\bibfnamefont {F.}~\bibnamefont {Henneberger}},\ }\bibfield  {title} {{\selectlanguage {en}\bibinfo {title} {Excitability of a {Semiconductor} {Laser} by a {Two}-{Mode} {Homoclinic} {Bifurcation}}},\ }\href {https://doi.org/10.1103/PhysRevLett.88.023901} {\bibfield  {journal} {\bibinfo  {journal} {Physical Review Letters}\ }\textbf {\bibinfo {volume} {88}},\ \bibinfo {pages} {023901} (\bibinfo {year} {2001})}\BibitemShut {NoStop}%
\bibitem [{\citenamefont {Liu}\ \emph {et~al.}(2017)\citenamefont {Liu}, \citenamefont {Li},\ and\ \citenamefont {Xiao}}]{liu_electromagnetically_2017}%
  \BibitemOpen
  \bibfield  {author} {\bibinfo {author} {\bibfnamefont {Y.-C.}\ \bibnamefont {Liu}}, \bibinfo {author} {\bibfnamefont {B.-B.}\ \bibnamefont {Li}},\ and\ \bibinfo {author} {\bibfnamefont {Y.-F.}\ \bibnamefont {Xiao}},\ }\bibfield  {title} {\bibinfo {title} {Electromagnetically induced transparency in optical microcavities},\ }\href {https://doi.org/10.1515/nanoph-2016-0168} {\bibfield  {journal} {\bibinfo  {journal} {Nanophotonics}\ }\textbf {\bibinfo {volume} {6}},\ \bibinfo {pages} {789} (\bibinfo {year} {2017})}\BibitemShut {NoStop}%
\bibitem [{\citenamefont {Hamel}\ \emph {et~al.}(2015)\citenamefont {Hamel}, \citenamefont {Haddadi}, \citenamefont {Raineri}, \citenamefont {Monnier}, \citenamefont {Beaudoin}, \citenamefont {Sagnes}, \citenamefont {Levenson},\ and\ \citenamefont {Yacomotti}}]{hamel_spontaneous_2015}%
  \BibitemOpen
  \bibfield  {author} {\bibinfo {author} {\bibfnamefont {P.}~\bibnamefont {Hamel}}, \bibinfo {author} {\bibfnamefont {S.}~\bibnamefont {Haddadi}}, \bibinfo {author} {\bibfnamefont {F.}~\bibnamefont {Raineri}}, \bibinfo {author} {\bibfnamefont {P.}~\bibnamefont {Monnier}}, \bibinfo {author} {\bibfnamefont {G.}~\bibnamefont {Beaudoin}}, \bibinfo {author} {\bibfnamefont {I.}~\bibnamefont {Sagnes}}, \bibinfo {author} {\bibfnamefont {A.}~\bibnamefont {Levenson}},\ and\ \bibinfo {author} {\bibfnamefont {A.~M.}\ \bibnamefont {Yacomotti}},\ }\bibfield  {title} {{\selectlanguage {en}\bibinfo {title} {Spontaneous mirror-symmetry breaking in coupled photonic-crystal nanolasers}},\ }\href {https://doi.org/10.1038/nphoton.2015.65} {\bibfield  {journal} {\bibinfo  {journal} {Nature Photonics}\ }\textbf {\bibinfo {volume} {9}},\ \bibinfo {pages} {311} (\bibinfo {year} {2015})}\BibitemShut {NoStop}%
\bibitem [{\citenamefont {Huang}\ \emph {et~al.}(2019)\citenamefont {Huang}, \citenamefont {Tran}, \citenamefont {Guo}, \citenamefont {Peters}, \citenamefont {Komljenovic}, \citenamefont {Malik}, \citenamefont {Morton},\ and\ \citenamefont {Bowers}}]{huang_high-power_2019}%
  \BibitemOpen
  \bibfield  {author} {\bibinfo {author} {\bibfnamefont {D.}~\bibnamefont {Huang}}, \bibinfo {author} {\bibfnamefont {M.~A.}\ \bibnamefont {Tran}}, \bibinfo {author} {\bibfnamefont {J.}~\bibnamefont {Guo}}, \bibinfo {author} {\bibfnamefont {J.}~\bibnamefont {Peters}}, \bibinfo {author} {\bibfnamefont {T.}~\bibnamefont {Komljenovic}}, \bibinfo {author} {\bibfnamefont {A.}~\bibnamefont {Malik}}, \bibinfo {author} {\bibfnamefont {P.~A.}\ \bibnamefont {Morton}},\ and\ \bibinfo {author} {\bibfnamefont {J.~E.}\ \bibnamefont {Bowers}},\ }\bibfield  {title} {{\selectlanguage {en}\bibinfo {title} {High-power sub-{kHz} linewidth lasers fully integrated on silicon}},\ }\href {https://doi.org/10.1364/OPTICA.6.000745} {\bibfield  {journal} {\bibinfo  {journal} {Optica}\ }\textbf {\bibinfo {volume} {6}},\ \bibinfo {pages} {745} (\bibinfo {year} {2019})}\BibitemShut {NoStop}%
\bibitem [{\citenamefont {Li}\ \emph {et~al.}(2023)\citenamefont {Li}, \citenamefont {Huang}, \citenamefont {Ma}, \citenamefont {Zhang}, \citenamefont {Xiao}, \citenamefont {Yang},\ and\ \citenamefont {Huang}}]{li_self-pulsing_2023}%
  \BibitemOpen
  \bibfield  {author} {\bibinfo {author} {\bibfnamefont {J.-C.}\ \bibnamefont {Li}}, \bibinfo {author} {\bibfnamefont {Y.-T.}\ \bibnamefont {Huang}}, \bibinfo {author} {\bibfnamefont {C.-G.}\ \bibnamefont {Ma}}, \bibinfo {author} {\bibfnamefont {Z.-N.}\ \bibnamefont {Zhang}}, \bibinfo {author} {\bibfnamefont {J.-L.}\ \bibnamefont {Xiao}}, \bibinfo {author} {\bibfnamefont {Y.-D.}\ \bibnamefont {Yang}},\ and\ \bibinfo {author} {\bibfnamefont {Y.-Z.}\ \bibnamefont {Huang}},\ }\bibfield  {title} {{\selectlanguage {en}\bibinfo {title} {Self-pulsing and dual-mode lasing in a square microcavity semiconductor laser}},\ }\href {https://doi.org/10.1364/OL.501029} {\bibfield  {journal} {\bibinfo  {journal} {Optics Letters}\ }\textbf {\bibinfo {volume} {48}},\ \bibinfo {pages} {4953} (\bibinfo {year} {2023})}\BibitemShut {NoStop}%
\end{thebibliography}%
\end{document}